\documentclass[11pt]{article}
\usepackage{cite}
\usepackage{amsmath,amsfonts,amssymb}
\usepackage[small,bf,hang]{caption}
\usepackage{slashed}

\def\hybrid{
        \topmargin -20pt
        \oddsidemargin 0pt
        \headheight 0pt \headsep 0pt
        \textwidth 6.25in 
        \textheight 9.5in 
        \marginparwidth .875in
        \parskip 5pt plus 1pt \jot = 1.5ex}

\hybrid

 \csname
@addtoreset\endcsname{equation}{section}

\def\moth{\mathsurround=0pt}
\newdimen\zo \zo=0pt

\def\tick{\leaders\hrule height 0.5ex depth 0pt \hskip 0.5pt}
\def\upboxfill{$\moth \setbox\zo\hbox{\tick}%
  \hskip 3pt\hbox to 0pt{$\tick$\hss}\hrulefill \hbox to 7.5pt{$\tick$\hss}$}

\def\dtick{\leaders\hrule height .34pt depth 0.5ex \hskip 0.5pt}
\def\downboxfill{$\moth \setbox\zo\hbox{\dtick}%
  \hskip 2pt\hbox to 0pt{$\dtick$\hss}\hrulefill \hbox to 2pt{$\dtick$\hss}$}

\def\bec{\begin{center}}
\def\ec{\end{center}}

 \def\det{{\rm det\,}}
\def\be{\begin{equation}}
\def\ee{\end{equation}}
\def\bea{\begin{eqnarray}}
\def\eea{\end{eqnarray}}
\def\ba{\begin{array}}
\def\ea{\end{array}}

\begin{document}

\begin{titlepage}
\rightline{}
\rightline{\tt MIT-CTP-4328}
\rightline{\tt  LMU-ASC 74/11}
\rightline{November 2011}
\begin{center}
\vskip 2.5cm
{\Large \bf {
${\cal N}=1$ Supersymmetric Double Field Theory
}}\\
\vskip 2.5cm
{\large {Olaf Hohm${}^1$ and Seung Ki Kwak${}^2$}}
\vskip 1cm
{\it {${}^1$Arnold Sommerfeld Center for Theoretical Physics}}\\
{\it {Theresienstrasse 37}}\\
{\it {D-80333 Munich, Germany}}\\
olaf.hohm@physik.uni-muenchen.de
\vskip 0.5cm
{\it {${}^2$Center for Theoretical Physics}}\\
{\it {Massachusetts Institute of Technology}}\\
{\it {Cambridge, MA 02139, USA}}\\
sk\_kwak@mit.edu
\vskip 0.7cm

\vskip 1.5cm
{\bf Abstract}
\end{center}

\vskip 0.5cm

\noindent
\begin{narrower}
We construct the ${\cal N}=1$ supersymmetric extension of double field theory for $D=10$,
including the coupling to an arbitrary number $n$ of abelian vector multiplets. This theory features
a local $O(1,9+n)\times O(1,9)$ tangent space symmetry under which the fermions transform.
It is shown that the supersymmetry transformations close into the generalized diffeomorphisms
of double field theory.

\end{narrower}

\vspace{4cm}

\end{titlepage}

\newpage

\section{Introduction}\setcounter{equation}{0}
Double field theory
is an approach to make the T-duality group $O(D,D)$  a manifest
symmetry of the massless sector of string theory
by doubling the $D$ space-time coordinates \cite{Hull:2009mi,Hull:2009zb,Hohm:2010jy,Hohm:2010pp}.
(See \cite{Siegel:1993th,Tseytlin:1990nb,Hohm:2010xe,Kwak:2010ew,
Hohm:2011ex,Hohm:2011zr,Hohm:2011cp,Hillmann:2009ci,Berman:2010is,West:2010ev,Jeon:2010rw,Jeon:2011cn,Jeon:2011kp,Jeon:2011vx,Schulz:2011ye,
Copland:2011yh,Thompson:2011uw,Albertsson:2011ux,Andriot:2011uh} for earlier work
and further developments.)
Thus, for $D=10$ the theory features a global $O(10,10)$
symmetry and depends formally on 20 coordinates,
but consistency requires an $O(10,10)$ 
invariant constraint
that locally removes the dependence on half of the coordinates.
Here we will construct the ${\cal N}=1$ supersymmetric
extension of double field theory for $D=10$.

Naively, one may suspect that such a construction is
impossible, for there simply are no supersymmetric theories
beyond eleven dimensions. The aforementioned constraint, however,
makes the supersymmetric extension feasible, because
for every solution of the constraint, locally the fields depend
only on ten coordinates.

The formulation of double field theory that is most useful
for our present purpose is the frame or vielbein formulation.
The double field theory can be written in terms of the
generalized metric
 \be\label{firstH}
  {\cal H}_{MN} \ = \  \begin{pmatrix}    g^{ij} & -g^{ik}b_{kj}\\[0.5ex]
  b_{ik}g^{kj} & g_{ij}-b_{ik}g^{kl}b_{lj}\end{pmatrix}\;,
 \ee
that takes values in $O(10,10)$ and combines the
space-time metric $g_{ij}$ and the Kalb-Ramond 2-form
$b_{ij}$. As usual, we may introduce frame fields $E_{M}{}^{A}$
and write 
 \be\label{framedef}
  {\cal H}_{MN} \ = \ E_{M}{}^{A}\,E_{N}{}^{B}\,\hat{\eta}_{AB}\;, \qquad
  \hat{\eta}_{AB} \ = \ \begin{pmatrix}   \eta_{ab}  &  0\\[0.5ex]
  0 & \eta_{\bar{a}\bar{b}}\end{pmatrix}\;,
 \ee
where $\eta$ denotes the standard Minkowski metric,
and we have split the flat or frame indices as $A=(a,\bar{a})$.
Consequently, in the frame formulation there is an
$O(1,9)_L\times O(1,9)_R$ `tangent space' gauge symmetry,
with $a,b\ldots=0,\ldots,9$ and $\bar{a},\bar{b}\ldots=0,\ldots,9$
denoting  $O(1,9)_L$ and $O(1,9)_R$ vector indices, respectively.
Such a frame formalism has been developed by Siegel
prior to the generalized metric
formulation \cite{Siegel:1993th}. Actually, Siegel's formalism allows also for the larger
tangent space group $GL(D)\times GL(D)$, but here we will restrict to the
Lorentz subgroups in order to be able to define the
corresponding spinor representations.
In this formalism one may introduce connections for the local
frame symmetry and construct invariant curvatures. This, in turn,
allows one to write an Einstein-Hilbert like action based on a
generalized curvature scalar ${\cal R}$, which provides an equivalent definition of
double field theory,
 \be\label{DFTaction}
  S \ = \ \int d^{10}x\, d^{10}\tilde{x}\,e^{-2d}\,{\cal R}(E,d)\;,
 \ee
where we defined
$e^{-2d}=\sqrt{g}e^{-2\phi}$.  In the frame formulation the theory
has a global $O(10,10)$ symmetry, a $O(1,9)_L\times O(1,9)_R$
gauge invariance and a `generalized diffeomorphism' symmetry.

In this paper we will introduce fermions that, as usual in
supergravity, are scalars under (generalized) diffeomorphisms
and $O(10,10)$, but which transform under the local tangent space
group $O(1,9)_L\times O(1,9)_R$.  The fermionic sector of supergravity
is thereby rewritten in a way that enlarges the local Lorentz group.
Similar attempts have in fact a long history, going back to the
work of de Wit and Nicolai in the mid 80's, in which they showed that
11-dimensional supergravity can be reformulated such that
it permits an enhanced tangent space symmetry \cite{deWit:1985iy}.
More recently, a very interesting paper appeared which showed in the context of
generalized geometry that type II supergravity can be reformulated such that
it permits a doubled Lorentz group \cite{Coimbra:2011nw}, as in double field theory, 
and our results are closely related
(see also \cite{Jeon:2011vx}).

We will introduce a gravitino field $\Psi_{a}$ that is a spinor
under $O(1,9)_R$ and a vector under $O(1,9)_L$, together with
a dilatino $\rho$, that is  a  spinor under $O(1,9)_R$.
The minimally supersymmetric extension of (\ref{DFTaction})
can then be written as
 \be\label{N=1DFTaction}
  S_{{\cal N}=1} \ = \ \int d^{10}x\, d^{10}\tilde{x}\,e^{-2d}\left({\cal R}(E,d)
  - \bar{\Psi}^{a}\gamma^{\bar{b}}\nabla_{\bar{b}}\Psi_{a}
 +\bar{\rho}\gamma^{\bar{a}}\nabla_{\bar{a}}\rho+2\bar{\Psi}^{a}\nabla_{a}\rho \right)\;.
 \ee
Here, the $\gamma^{\bar{a}}$ are ten-dimensional gamma matrices,
which have to be thought of as gamma matrices of $O(1,9)_R$, so that
all suppressed spinor indices in (\ref{DFTaction}) are $O(1,9)_R$
spinor indices. Moreover, the covariant derivatives $\nabla$ are with respect
to the connections introduced by Siegel \cite{Siegel:1993th}, and
therefore the action is manifestly $O(1,9)_L\times O(1,9)_R$
invariant.

We will show that (\ref{N=1DFTaction}), up to field redefinitions,
reduces precisely to the standard minimal ${\cal N}=1$ action
in ten dimensions. In this paper we will not consider higher-order 
fermi terms.  Formally, (\ref{N=1DFTaction})
is contained in the results of \cite{Coimbra:2011nw} through the
straightforward truncation from ${\cal N}=2$ to ${\cal N}=1$.
The main difference between generalized geometry, which
was the starting point in \cite{Coimbra:2011nw}, and
double field theory is that in the former the coordinates are
not doubled but only the tangent space. Consequently,
in generalized geometry only the tangent space symmetry
is enhanced, while double field theory features also a
global $O(D,D)$ symmetry. With the fermions being singlets
under $O(D,D)$, this symmetry is somewhat trivially realized
on the fermionic sector, and therefore our results
for the minimal ${\cal N}=1$ theory 
are largely contained in those of generalized geometry
given in \cite{Coimbra:2011nw}.
In the spirit of double field theory, however, it is reassuring 
to verify
closure of the supersymmetry transformations into
generalized diffeomorphisms
and supersymmetric invariance of (\ref{N=1DFTaction}),
both modulo the $O(D,D)$ invariant constraint.
This will be done in sec.~2 of this paper.

As the main new result, we will present in sec.~3 the
double field theory extension of ${\cal N}=1$
supergravity in $D=10$ coupled to an arbitrary
number $n$ of (abelian) vector multiplets.
For $n=16$ this is the low-energy effective action
of heterotic superstring theory truncated
to the Cartan subalgebra of $SO(32)$ or $E_8\times E_8$.
As has been shown in \cite{Hohm:2011ex},
the coupling of gauge vectors $A_{i}{}^{\alpha}$
can be neatly described
by enlarging the generalized metric
(\ref{firstH}) to an $O(10+n,10)$ matrix that
naturally contains the $A_{i}{}^{\alpha}$.
In the frame formulation this theory features,
in addition, a $O(1,9+n)\times O(1,9)$
tangent space symmetry. The fermionic fields
will still be spinors under $O(1,9)$, but
$\Psi_{a}$ is now a vector under $O(1,9+n)$.
Remarkably, it turns out that the same action (\ref{N=1DFTaction}),
but written with respect to these enlarged fields,
reproduces precisely the ${\cal N}=1$ supergravity
coupled to abelian vector multiplets, with the
gauginos originating from the additional
components of the $\Psi_{a}$.

Let us finally mention that in the work of Siegel the construction
proceeds immediately in ${\cal N}=1$ superspace \cite{Siegel:1993th}.
Therefore, our results on the ${\cal N}=1$ theory, including the coupling to vector multiplets, must be related to the
construction of Siegel, but we have not been
daring enough to attempt an explicit verification.

\textit{Note added:} After the submission of the first version of this 
paper to the arxiv, \cite{Jeon:2011sq} appeared, which overlaps 
with our section 2.

\section{Minimal ${\cal N}=1$ Double Field Theory for $D=10$}
In this section we introduce the minimal ${\cal N}=1$ theory. First,
we review the vielbein formalism with local  $O(1,9)_L\times O(1,9)_R$
symmetry. Second, we introduce the ${\cal N}=1$ double field theory and
prove its supersymmetric invariance. In the third subsection we verify
that it reduces to conventional ${\cal N}=1$
supergravity upon setting the new derivatives to zero.

\subsection{Vielbein formulation with local $O(1,9)\times O(1,9)$ symmetry}\label{framesec}
We start by reviewing some generalities on the vielbein formulation
of double field theory, which is contained in Siegel's frame formalism \cite{Siegel:1993th}.
We refer to \cite{Hohm:2010xe} for a self-contained presentation of this
formulation.  
The fundamental bosonic fields are the frame field $E_{A}{}^{M}$
and the dilaton $d$ that
depend both on doubled coordinates $X^{M}= (\tilde{x}_i, x^i)$.
The frame field is subject to
local $O(1,9)_L\times O(1,9)_R$ transformations acting on the
index $A=(a,\bar{a})$ and global $O(10,10)$ transformations acting
on the index $M$, which read infinitesimally
 \be
  \delta E_{A}{}^{M} \ = \ k^{M}{}_{N}  E_{A}{}^{N}+\Lambda_{A}{}^{B}(X)E_{B}{}^{M}\,,
  \qquad
  k\ \in \ \frak{o}(10,10)\;, \quad \Lambda(X) \ \in \ \frak{o}(1,9)_L \oplus \frak{o}(1,9)_R\;, 
 \ee
where the parameters take values in the respective Lie algebras.  
The double field theory is invariant under a
`generalized diffeomorphism' symmetry parameterized by
$\xi^{M}=(\tilde{\xi}_i,\xi^{i})$ that combines the $b$-field
1-form gauge parameter $\tilde{\xi}_i$ with the vector-valued
diffeomorphism parameter $\xi^{i}$,
 \be\label{gendiff}
  \delta_{\xi}E_{A}{}^{M} \ = \ \widehat{\cal L}_{\xi}E_{A}{}^{M}
  \ \equiv \ \xi^{N}\partial_{N}E_{A}{}^{M}
  +\big(\partial^{M}\xi_{N}-\partial_{N}\xi^{M}\big)E_{A}{}^{N}\;.
 \ee
Here, $\partial_M=(\tilde{\partial}^i,\partial_{i})$ are the
doubled partial derivatives. The right-hand side of (\ref{gendiff})
defines a generalized Lie derivative that can similarly be defined
for an $O(D,D)$ tensor with an arbitrary number of upper and
lower indices. On the dilaton $d$ these gauge transformations read
 \be
  \delta_{\xi}d \ = \ \xi^M\partial_M d-\frac{1}{2}\partial_M \xi^{M}\;.
 \ee  
The gauge transformations close and leave the action invariant
modulo the `strong constraint'
 \be\label{strongconstr}
  \eta^{MN}\partial_{M}\partial_{N} \ = \ 0\;, \qquad
  \eta^{MN} \ = \ \begin{pmatrix}    0 & {\bf 1} \\[0.5ex]
  {\bf 1} & 0 \end{pmatrix}\;,
 \ee
when acting on arbitrary fields and parameters and all their products.
Here, $\eta_{MN}$ denotes the $O(10,10)$ invariant metric, which
will be used to raise and lower $O(10,10)$ indices.
This constraint implies that locally all fields depend only on half of
the coordinates, for instance only on the $x^i$.

We have to impose covariant constraints on the frame field
in order to describe only the physical degrees of freedom.
These constraints are written in terms of the tangent space metric
 \be
  {\cal G}_{AB} \ \equiv \ E_{A}{}^{M}\,E_{B}{}^{N}\,\eta_{MN}\;,
 \ee
resulting from the $O(10,10)$ invariant metric $\eta$,
and which will be used to raise and lower flat indices.
We require the $O(1,9)_L\times O(1,9)_R$ covariant constraints
 \be\label{Gconstr}
  {\cal G}_{a\bar{b}} \ = \ 0\;, \qquad
  {\cal G}_{ab} \ = \ \eta_{ab}\;, \qquad
  {\cal G}_{\bar{a}\bar{b}} \ = \ -\eta_{\bar{a}\bar{b}}\;.
 \ee
Note that the relative minus sign entering here is necessary due to
the $(10,10)$ signature of ${\cal G}_{AB}$. It is a matter of convention
to which metric we assign the minus sign, but once the choice
is made the symmetry between unbarred and barred indices is
broken. Since flat indices are raised and lowered
with ${\cal G}_{AB}$, (\ref{Gconstr}) leads to some unconventional signs
when comparing below to standard expressions for, say, the spin connection.
We will comment on this in due course.

A particular solution of these constraints, giving
rise to the generalized metric (\ref{firstH})
according to (\ref{framedef}), is given by
  \be\label{coset}
   E_{A}{}^{M} \ = \
    \begin{pmatrix} E_{ai} & E_{a}{}^{i} \\
    E_{\bar{a}i} & E_{\bar{a}}{}^{i} \end{pmatrix} \ = \
   \frac{1}{\sqrt{2}} \begin{pmatrix} e_{ia}+b_{ij}e_{a}{}^{j} & e_{a}{}^{i} \\
   -e_{i\bar{a}}+b_{ij}e_{\bar{a}}{}^{j} & e_{\bar{a}}{}^{i} \end{pmatrix}\;,
 \ee
where $e$ is the vielbein of the conventional metric, $g=e\,\eta\,e^{T}$.
We stress that when writing (\ref{coset}) the tangent space symmetry
is gauge-fixed to the diagonal subgroup of $O(1,9)_L\times O(1,9)_R$, as
is clear from the fact that $e$ carries in (\ref{coset}) both unbarred
and barred indices. In order to define the supersymmetric double field theory,
however, (\ref{coset}) is never used. Rather, we view
the (constrained) vielbein $E_{A}{}^{M}$ as the fundamental field and so the construction is manifestly
invariant under two copies of the local Lorentz group. It is only when comparing
to the standard formulation of supergravity that we have to use (\ref{coset})
and to partially gauge-fix.\footnote{This differs from the construction
in \cite{Coimbra:2011nw} and \cite{Jeon:2011cn,Jeon:2011vx}, where two independent vielbein fields are introduced,
one transforming under $O(1,9)_L$ and one transforming under $O(1,9)_R$.}

Let us now turn to the definition of connections and covariant derivatives.
We first note that the partial derivative of a field $S$ that transforms as a scalar
under $\xi^{M}$, i.e.,
 \be
  \delta_{\xi}S \ = \ \xi^{M}\partial_{M}S\;,
 \ee
transforms covariantly with a generalized Lie derivative \cite{Hohm:2010xe}.
This does not hold for higher tensors, which in turn necessitates
the introduction of covariant derivatives.
Given the frame field $E_{A}{}^{M}$, we introduce the
`flattened' partial derivative\footnote{Here we introduced
a factor of $\sqrt{2}$ for later convenience. With the constraints
on the connections to be imposed below, the covariant derivatives
$\nabla_{A}$ given here are $\sqrt{2}$ times the covariant
derivatives in \cite{Hohm:2010xe}.}
 \be
  E_{A} \ \equiv \ \sqrt{2}E_{A}{}^{M}\partial_{M}\;.
 \ee
We can then introduce $O(1,9)_L\times O(1,9)_R$
covariant derivatives
  \be\label{covder}
  \nabla_{A}V_B \ = \ E_{A}V_{B}+\omega_{AB}{}^{C}V_{C}\;, \qquad
  \nabla_{A}V^{B} \ = \ E_{A}V^{B}-\omega_{AC}{}^{B}V^{C}\;,
 \ee
where we stress that the only non-trivial connections
are $\omega_{Ab}{}^{c}$ and $\omega_{A\bar{b}}{}^{\bar{c}}$.

Next, we briefly summarize which connection components
can be determined in terms of $E_{A}{}^{M}$ and $d$ upon imposing
covariant constraints. First, in order to be compatible with
the constancy of the tangent space metric ${\cal G}_{AB}$, the
symmetric part $\omega_{A(BC)}$, where indices have been lowered
with ${\cal G}$, is zero. Thus, $\omega_{ABC}$ is antisymmetric
in its last two indices. Second, we can impose a generalized torsion
constraint, which  reads
 \be\label{NewTorsion}
  {\cal T}_{ABC} \ \equiv \ \Omega_{ABC} + 3 \omega_{[ABC]} 
  \ = \ 0 \; ,
 \ee
where we introduced  the `generalized coefficients of anholonomy'
 \be\label{newanhol}
  \Omega_{ABC} \ = \ 3f_{[ABC]}\;, \qquad
  f_{ABC} \ \equiv \ (E_{A}E_{B}{}^{M})E_{CM}\;.
 \ee
We note that $f_{ABC}$ is antisymmetric in its last two indices as a 
consequence of the constancy of ${\cal G}_{AB}$. 
Specializing the constraint (\ref{NewTorsion}) to ${\cal T}_{a\bar{b}\bar{c}}=0$
and ${\cal T}_{\bar{a}bc}=0$, we derive the 
following solution for the `off-diagonal' components
 \be\label{Omcomp}
  \omega_{a\bar{b}\bar{c}} \ = \ -\Omega_{a\bar{b}\bar{c}}\;, \qquad
  \omega_{\bar{a}bc} \ = \ -\Omega_{\bar{a}bc}\;.
 \ee
For later use let us determine these connection components
for the gauge choice (\ref{coset}) of the frame field,
setting $\tilde{\partial}^i=0$. We compute
with (\ref{newanhol})
 \be
  f_{a\bar{b}\bar{c}} \ = \ e_{a}{}^{i}e_{[\bar{b}}{}^{j}\partial_{i}
  e_{j\bar{c}]}+\frac{1}{2}e_{a}{}^{i}e_{\bar{b}}{}^{k}e_{\bar{c}}{}^{j}\partial_i b_{jk}\;, \qquad
  f_{\bar{b}a\bar{c}} \ = \ e_{\bar{b}}{}^{i}e_{(a}{}^{j}\partial_{i}
  e_{j\bar{c})}+\frac{1}{2}e_{\bar{b}}{}^{i}e_{a}{}^{k}e_{\bar{c}}{}^{j}\partial_i b_{jk}\;,
 \ee
from which we derive
 \be\label{finaloff}
  \omega_{a\bar{b}\bar{c}} \ = \ -\omega_{a\bar{b}\bar{c}}^{\rm L}(e)+\frac{1}{2}e_{a}{}^{i}e_{\bar{b}}{}^{j}e_{\bar{c}}{}^{k}
  H_{ijk}\;,
 \ee
where $\omega^{\rm L}$ denotes the standard Levi-Civita spin connection
expressed in terms of the vielbein,
 \be\label{standardspin}
  \omega_{a\bar{b}\bar{c}}^{\rm L}(e) \ = \ e_{[a}{}^{i}e_{\bar{b}]}{}^{j}\partial_{i}e_{j\bar{c}}
  -e_{[\bar{b}}{}^{i}e_{\bar{c}]}{}^{j}\partial_{i}e_{ja}
  +e_{[\bar{c}}{}^{i}e_{a]}{}^{j}\partial_{i}e_{j\bar{b}}\;.
 \ee
Similarly, one finds
 \be\label{finaloffother}
   \omega_{\bar{a}bc} \ = \ \omega^{\rm L}_{\bar{a}bc}(e)+\frac{1}{2}H_{\bar{a}bc}\;,
 \ee
where  we flattened the indices of $H$ as in (\ref{finaloff}).\footnote{We note that the
relative sign between $\omega_{a\bar{b}\bar{c}}$ and $\omega_{a\bar{b}\bar{c}}^{\rm L}$ in (\ref{finaloff})
is due to the fact that we lower barred indices with ${\cal G}_{\bar{a}\bar{b}}=-\eta_{\bar{a}\bar{b}}$, see eq.~(\ref{Gconstr}),
while in the standard expression (\ref{standardspin}) for the spin connection
the index is lowered with $\eta_{\bar{a}\bar{b}}$.
Correspondingly, there is no relative sign in (\ref{finaloffother}) because here indices
are lowered with ${\cal G}_{ab}=\eta_{ab}$.}

For the `diagonal' components, having either only unbarred or barred
indices, the totally antisymmetric parts are determined by (\ref{NewTorsion}) as follows
 \be\label{antisol}
  \omega_{[abc]} \ = \ -\frac{1}{3}\Omega_{[abc]} \ = \ -f_{[abc]} \;, \qquad
  \omega_{[\bar{a}\bar{b}\bar{c}]} \ = \
  -\frac{1}{3}\Omega_{[\bar{a}\bar{b}\bar{c}]} \ = \ -f_{[\bar{a}\bar{b}\bar{c}]}\;.
 \ee
Again, we may determine these connections for the gauge choice (\ref{coset})
and $\tilde{\partial}^i=0$. One finds,
\be \label{screl}
\omega_{[a b c]} \ = \ \omega^{\rm L}_{[a b c]}(e) + \frac{1}{6} H_{abc} \;  , \qquad
\omega_{[\bar{a} \bar{b} \bar{c}]} \ = \ -\omega^{\rm L}_{[\bar{a} \bar{b} \bar{c}]}(e)
+ \frac{1}{6} H_{\bar{a} \bar{b} \bar{c} } \; ,
\ee
where we flattened the indices on $H$.

The torsion constraint leaves the mixed Young tableaux representation
in $\omega_{abc}$ and $\omega_{\bar{a}\bar{b}\bar{c}}$ undetermined,
but its trace part can be fixed by imposing a covariant constraint
that allows for partial integration in presence of the dilaton density,
 \be\label{dilconstr}
  \int e^{-2d}\, V\nabla_{A}V^{A} \ = \ -\int e^{-2d}\,V^{A}\nabla_{A}V \;,
 \ee
for arbitrary $V$ and $V^{A}$. This implies
  \be\label{tracepart}
  \omega_{BA}{}^{B} \ = \ -\tilde{\Omega}_{A} \ \equiv \ -\sqrt{2}e^{2d}\partial_{M}\big(E_{A}{}^{M}e^{-2d}\big)\;,
 \ee
where we introduced $\tilde{\Omega}_{A}$ for later use.
Note that this determines precisely $\omega_{ba}{}^{b}$ and $\omega_{\bar{b}\bar{a}}{}^{\bar{b}}$, because
the last two indices cannot be mixed.

Finally, we can introduce an invariant scalar curvature and Ricci tensor.
In the frame formalism there is an invariant curvature tensor ${\cal R}_{ABCD}$,
but it is generally not a function of the determined connections only.
For the derived curvature scalar and Ricci tensor, however, it depends
only on the determined connections.
Without repeating the details of the construction, we give the
explicit expressions.

The scalar curvature can be defined as the trace over, say, barred indices as follows
 \be\label{scalarcurv}
 \begin{split}
  {\cal R} \ &\equiv \  -{\cal R}_{\bar{a} \bar{b}}{}^{\bar{a} \bar{b}} \ = \
  -2 E_{\bar{a}} \omega_{\bar{b}}{}^{\bar{a} \bar{b}} - \frac{3}{2} \omega_{[\bar{a} \bar{b} \bar{c} ]}\,  \omega^{[\bar{a} \bar{b} \bar{c} ]}
  +  \omega_{\bar{a}}{}^{\bar{c} \bar{a}}\, \omega_{\bar{b} \bar{c}}{}^{\bar{b}} - \frac{1}{2} \omega_{a \bar{b} \bar{c}}\, \omega^{a \bar{b} \bar{c}}  \\
   \ &= \ 2  E_{\bar{a}} \tilde{\Omega}^{\bar{a}} + \tilde{\Omega}_{\bar{a}}^2 - \frac{1}{2} \Omega_{\bar{a} \bar{b} c}{}^{2}
   - \frac{1}{6}\Omega_{[\bar{a} \bar{b} \bar{c}]}{}^{2}\;,
 \end{split}
 \ee
where we have written in the second line the explicit expression in terms of $\Omega$ and thereby
in terms of the physical fields.
The Ricci tensor reads
 \be
  {\cal R}_{a \bar{b}} \ = \
  E_{\bar{c}} \omega_{a \bar{b}}{}^{\bar{c}} - E_{a} \omega_{\bar{c} \bar{b} }{}^{\bar{c}} + \omega_{d \bar{b} }{}^{\bar{c}}\, \omega_{\bar{c} a} {}^{d}
  - \omega_{a \bar{b}}{}^{ \bar{d} }\, \omega_{\bar{c} \bar{d}}{}^{\bar{c}} \;.
  \ee
These curvature invariants can be obtained by variation of the (bosonic) double field theory action. 
In order to see 
this it is convenient to introduce the variation 
\be\label{Delta}
  \Delta E_{AB} \ := \ E_{B}{}^{M}\delta E_{A M}\;, 
 \ee
which is antisymmetric in $A,B$ as a consequence of the constancy of ${\cal G}_{AB}$.  
Under the local $O(1,9)_L\times O(1,9)_R$ this variation reads $\Delta E_{ab}=\Lambda_{ab}$
and $\Delta E_{\bar{a}\bar{b}}=\Lambda_{\bar{a}\bar{b}}$.  
Thus, only the off-diagonal variation is not pure-gauge and the corresponding   
general variation of the action (\ref{DFTaction})
can be written in terms of the curvatures as \cite{Hohm:2010xe}
 \be\label{GenVar}
   \delta S \ = \ -2\int dxd\tilde{x}\,e^{-2d}\left(\delta d\,{\cal R}+
   \Delta E_{a\bar{b}}\,{\cal R}^{a\bar{b}}\right)\;,
 \ee
which will be used below.

\subsection{${\cal N}=1$ Double Field Theory}
We give now the ${\cal N}=1$ supersymmetric extension of double field theory
in the frame formulation reviewed above. The fermionic fields are the `gravitino'
$\psi_a$ and the `dilatino` $\rho$, and we will later see how they are related
to the conventional gravitino and dilatino via a field redefinition.
These fields are scalars under $O(10,10)$ and generalized diffeomorphisms and,
together with the ${\cal N}=1$ supersymmetry parameter $\epsilon$,
transform under the local $O(1,9)_L\times O(1,9)_R$ as follows
 \be\label{assigment}
 \begin{split}
  \Psi_{a}\;&:\qquad \text{vector of}\;\;O(1,9)_{L}\;,\;\; \text{spinor of}\;\; O(1,9)_{R}\;, \\
  \rho\;&:\qquad \text{spinor of}\;\; O(1,9)_{R}\;, \\
  \epsilon\;&:\qquad \text{spinor of}\;\; O(1,9)_{R}\;.
 \end{split}
 \ee
The ${\cal N}=1$ supersymmetric extension of (\ref{DFTaction}) is given by (\ref{N=1DFTaction}),
 \be\label{N=1DFTactionMain}
  S_{{\cal N}=1} \ = \ \int dx d\tilde{x}\,e^{-2d}\left({\cal R}(E,d)
  - \bar{\Psi}^{a}\gamma^{\bar{b}}\nabla_{\bar{b}}\Psi_{a}
 +\bar{\rho}\gamma^{\bar{a}}\nabla_{\bar{a}}\rho+2\bar{\Psi}^{a}\nabla_{a}\rho \right)\;,
 \ee
where all covariant derivatives are with respect to the connections introduced above.
We will see below that in here and in the supersymmetry rules all undetermined connections drop out. 
When acting on $O(1,9)_R$ spinors the covariant derivatives are given by
 \be\label{covder}
  \nabla_{a} \ = \ E_{a}-\frac{1}{4}\omega_{a\bar{b}\bar{c}}\gamma^{\bar{b}\bar{c}}\;, \qquad
  \nabla_{\bar{a}} \ = \ E_{\bar{a}}-\frac{1}{4}\omega_{\bar{a}\bar{b}\bar{c}}\gamma^{\bar{b}\bar{c}}\;.
 \ee
We observe that (\ref{N=1DFTactionMain}) is manifestly $O(1,9)_L\times O(1,9)_R$ invariant,
because unbarred and barred indices are properly contracted, and the
$\gamma^{\bar{a}}$ are gamma matrices of $O(1,9)_R$, 
so that all suppressed spinor indices belong to $O(1,9)_R$.
More precisely, we define the $\gamma^{\bar{a}}$ to satisfy 
 \be\label{CLiff}
  \big\{\gamma^{\bar{a}},\gamma^{\bar{b}}\big\} \ = \ -2{\cal G}^{\bar{a}\bar{b}} \ = \ 2\eta^{\bar{a}\bar{b}}\;, 
 \ee 
where the signs are such that the $\gamma^{\bar{a}}$ can be 
chosen to be conventional gamma matrices in ten dimensions. 
We note that, according to our convention, on $\gamma_{\bar{a}}$ 
the index is lowered with ${\cal G}_{\bar{a}\bar{b}}=-\eta_{\bar{a}\bar{b}}$ so that 
it differs from the conventional ten-dimensional gamma matrix with a lower index 
by a sign. Similarly, the minus signs in
(\ref{covder}) are due to the lowering of indices on $\omega_{A\bar{b}\bar{c}}$ with ${\cal G}_{\bar{a}\bar{b}}$.
Let us finally stress that the assignment (\ref{assigment}) of $O(1,9)_L \times O(1,9)_R$ representations
is related to the constraint (\ref{Gconstr}). We could have chosen the opposite signatures for
${\cal G}_{ab}$ and ${\cal G}_{\bar{a}\bar{b}}$, 
but then supersymmetry would require the gravitino to be a vector
under  $O(1,9)_R$ and a spinor under $O(1,9)_L$.

The action (\ref{N=1DFTactionMain}) is manifestly invariant under
generalized diffeomorphisms, 
 \be
 \begin{split}
    \delta_{\xi}E_{A}{}^{M} \ &= \ \widehat{\cal L}_{\xi}E_{A}{}^{M} \; ,   \qquad\;\; \,   \delta_{\xi}d \ = \ \xi^M\partial_M d-\frac{1}{2}\partial_M \xi^{M}\;, \\
   \delta_{\xi}\Psi_{a} \ &= \ \xi^{M}\partial_{M}\Psi_{a}\;,  \qquad
   \delta_{\xi}\rho \ = \ \xi^{M}\partial_{M}\rho\;,
 \end{split}
 \ee
because with the fermions transforming as scalars the (flattened) derivatives in (\ref{N=1DFTactionMain}) transform covariantly.
In addition, the action is invariant under the ${\cal N}=1$
supersymmetry transformations \cite{Coimbra:2011nw}
 \be\label{DFTSUSY}
 \begin{split}
   \Delta^{}_{\epsilon} E^{}_{a\bar{b}} \ &= \ -\frac{1}{2} \bar{\epsilon}\,\gamma_{\bar{b}}\Psi^{}_{a}\;, \qquad \;
   \delta_{\epsilon} d \ = \  -  \frac{1}{4} \bar{\epsilon} \rho\;, \\
   \delta_{\epsilon} \Psi_{a} \ &= \ \nabla_a \epsilon \;, \qquad \qquad \;\;\;
   \delta_{\epsilon} \rho \ = \   \gamma^{\bar{a}} \nabla_{\bar{a}} \epsilon \;. 
 \end{split}
 \ee
Here, we have written the transformation of the frame field
in terms of the variation (\ref{Delta}). Due to the $O(1,9)_L\times O(1,9)_R$ 
gauge freedom, we can assume for the diagonal supersymmetry variations 
$\Delta_{\epsilon}E_{ab}=\Delta_{\epsilon}E_{\bar{a}\bar{b}}=0$.

Let us now verify that (\ref{N=1DFTactionMain}) is invariant under (\ref{DFTSUSY}), again 
up to higher-order fermi terms. 
We start with the variation of the bosonic part, which can be obtained directly by
inserting the fermionic supersymmetry rules of (\ref{DFTSUSY}) into (\ref{GenVar}),
 \be\label{bosvar}
  e^{2d}\,\delta_{\epsilon}{\cal L}_{\rm B} \ = \ \frac{1}{2}\bar{\epsilon}\rho{\cal R} + \bar{\epsilon}
  \gamma_{\bar{b}}\Psi_a {\cal R}^{a\bar{b}}\;,
 \ee
where we denoted the bosonic Lagrangian by ${\cal L}_{\rm B}$.
Denoting the fermionic part similarly by ${\cal L}_{\rm F}$, one finds
 \be\label{fermvar}
 \begin{split}
  e^{2d}\, \delta_{\epsilon}{\cal L}_{\rm F} \ &= \ - 2    \bar{\Psi}^{a}\gamma^{\bar{b}}\nabla_{\bar{b}}  \nabla_a \epsilon
  +2  \bar{\rho} \gamma^{\bar{a}  } \nabla_{\bar{a} } \big(\gamma^{\bar{b} } \nabla_{\bar{b} }  \epsilon\big)
  +2\nabla^{a}\bar{\epsilon}\,\nabla_{a}\rho+2     \bar{\Psi}^{a}  \nabla_{a} \big( \gamma^{\bar{b}}\nabla_{\bar{b}} \epsilon\big) \\
  \ &= \ - 2   \bar{\Psi}^{a} \big[  \gamma^{\bar{b}}  \nabla_{\bar{b}}  ,  \nabla_{a} \big] \epsilon
  + 2  \bar{\rho} \left( \gamma^{\bar{a}  } \nabla_{\bar{a} } \gamma^{\bar{b} } \nabla_{\bar{b} }-  \nabla^{a} \nabla_{a}  \right) \epsilon\;.
 \end{split}
 \ee
Here we have used that according to (\ref{dilconstr})
the covariant derivatives allow us to freely partially integrate
in presence of the dilaton density. Moreover, in the second line we have combined the
first and last and the second and third term. We can now use the identities \cite{Coimbra:2011nw}
 \be\label{CurvIdent}
 \begin{split}
    \left( \gamma^{\bar{a}  } \nabla_{\bar{a} } \gamma^{\bar{b} } \nabla_{\bar{b} }-  \nabla^{a} \nabla_{a}  \right) \epsilon \ &= \
   - \frac{1}{4}{\cal R}\epsilon\;, \\
   \Big[  \gamma^{\bar{b}}  \nabla_{\bar{b}}  ,  \nabla_{a} \Big] \epsilon \ &= \ - \frac{1}{2} \gamma^{\bar{b}}{\cal R}_{a\bar{b}}\epsilon\;,
  \end{split}
 \ee
which will be proved in the appendix, to see that this cancels precisely the variation (\ref{bosvar}) of the bosonic term,
proving supersymmetric invariance.

We turn now to the closure of the supersymmetry transformations. 
Since these are an invariance of the action (\ref{N=1DFTactionMain})
they must close into the other local symmetries of the theory, which 
are generalized diffeomorphisms and the doubled local Lorentz 
transformations $O(1,9)_L\times O(1,9)_R$. It is instructive, however, 
to investigate this explicitly, and so we verify in the following 
closure on the bosonic fields.   
For the dilaton we compute
 \be
  \big[\delta_{\epsilon_1},\delta_{\epsilon_2}\big]d \ = \ \frac{1}{4}\big(\bar{\epsilon}_1\gamma^{\bar{a}}\nabla_{\bar{a}}\epsilon_2-
  \bar{\epsilon}_2\gamma^{\bar{a}}\nabla_{\bar{a}}\epsilon_1\big) \ = \
  \frac{1}{4}\bar{\epsilon}_1\gamma^{\bar{a}}\big(E_{\bar{a}}-\frac{1}{4}\omega_{\bar{a}\bar{b}\bar{c}}\gamma^{\bar{b}\bar{c}}\big)\epsilon_2
  -(1 \leftrightarrow 2)\;.
 \ee
Let us work out the first term in here,
 \be\label{step123}
  \frac{1}{4}\bar{\epsilon}_1\gamma^{\bar{a}}E_{\bar{a}}\epsilon_2-(1 \leftrightarrow 2) \ = \
  \frac{\sqrt{2}}{4}\bar{\epsilon}_1\gamma^{\bar{a}}E_{\bar{a}}{}^{M}\partial_{M}\epsilon_2-(1 \leftrightarrow 2)
  \ = \ \frac{1}{2\sqrt{2}}E_{\bar{a}}{}^{M}\partial_{M}\big(\bar{\epsilon}_1\gamma^{\bar{a}}\epsilon_2\big)\;,
 \ee
using  $\bar{\epsilon}_1\gamma^{\bar{a}}\epsilon_2=-\bar{\epsilon}_2\gamma^{\bar{a}}\epsilon_1$.
For the second term we compute
 \be
 - \frac{1}{16}\omega_{\bar{a}\bar{b}\bar{c}}\bar{\epsilon}_1\gamma^{\bar{a}}\gamma^{\bar{b}\bar{c}}\epsilon_2-(1 \leftrightarrow 2)
  \ = \  - \frac{1}{16}\omega_{\bar{a}\bar{b}\bar{c}}\bar{\epsilon}_1\big(\gamma^{\bar{a}\bar{b}\bar{c}}-2{\cal G}^{\bar{a}[\bar{b}}\gamma^{\bar{c}]}\big)\epsilon_2-(1 \leftrightarrow 2)\;.
 \ee
The first term in here vanishes due to the antisymmetrization in $(1 \leftrightarrow 2)$ and
$\bar{\epsilon}_1\gamma^{\bar{a}\bar{b}\bar{c}}\epsilon_2=\bar{\epsilon}_2\gamma^{\bar{a}\bar{b}\bar{c}}\epsilon_1$.
The second term gives with (\ref{tracepart})
 \be
 - \frac{1}{4}  \omega_{\bar{a}\bar{c}}{}^{\bar{a}}\bar{\epsilon}_1\gamma^{\bar{c}}\epsilon_2 \ = \
   \frac{1}{2\sqrt{2}}  \big(\partial_{M}E_{\bar{c}}{}^{M}-2E_{\bar{c}}{}^{M}\partial_{M}d  \big)\bar{\epsilon}_1\gamma^{\bar{c}}\epsilon_2 \;.
 \ee
The first term in here combines with (\ref{step123}) to give $\tfrac{1}{2\sqrt{2}}\partial_{M}(E_{\bar{c}}{}^{M}\bar{\epsilon}_1\gamma^{\bar{c}}\epsilon_2)$.
The second term takes the form of a transport term so that we have shown in total
 \be\label{dclosure}
  \big[\delta_{\epsilon_1},\delta_{\epsilon_2}\big]d \ = \  \xi^{M}\partial_{M}d-\frac{1}{2}\partial_{M}\xi^{M}\;, \qquad
  \xi^{M} \ = \ -\frac{1}{\sqrt{2}}E_{\bar{a}}{}^{M}\,\bar{\epsilon}_1\gamma^{\bar{a}}\epsilon_2\;.
 \ee
Thus, the supersymmetry transformations close into generalized diffeomorphisms, as required.

Next, we verify closure on $E_{A}{}^{M}$.
We compute 
 \be\label{Eclosure}
 \begin{split}
  \big[ \delta_{\epsilon_1},\delta_{\epsilon_2}\big] E_{aM} \ &= \ \delta_{\epsilon_1}\big(E_{M}{}^{B}E_{B}{}^{N}\delta_{\epsilon_2}E_{aN}\big)-(1 \leftrightarrow 2) \\
  \ &= \ \delta_{\epsilon_1}\big(E_{M}{}^{\bar{b}}\Delta_{\epsilon_2}E_{a\bar{b}}\big)-(1 \leftrightarrow 2) 
  \ = \ -\frac{1}{2}\delta_{\epsilon_1}\big(E_{M}{}^{\bar{c}}\bar{\epsilon}_2\gamma_{\bar{c}}\Psi_{a}\big) -(1 \leftrightarrow 2)     \;,
 \end{split}
 \ee
where we used that we can set $\Delta_{\epsilon} E_{ab}=0$ by an appropriate $O(1,9)_L$ transformation, 
and we relabeled an index in the last equality. 
In order to disentangle the generalized diffeomorphisms and local  $O(1,9)_L\times O(1,9)_R$ transformations 
we project (\ref{Eclosure}) by multiplying with $E_{\bar{b}}{}^{M}$ and $E_{b}{}^{M}$, respectively. 
For the first we obtain 
 \be
 \begin{split}
  E_{\bar{b}}{}^{M}\big[ \delta_{\epsilon_1},\delta_{\epsilon_2}\big] E_{aM} \ &= \ -\frac{1}{2}E_{\bar{b}}{}^{M}E_{M}{}^{\bar{c}}\bar{\epsilon}_{2}\gamma_{\bar{c}}
  \nabla_{a}\epsilon_1 -(1 \leftrightarrow 2)   \\
  \ &= \ -\frac{1}{2}\bar{\epsilon}_2\gamma_{\bar{b}}\big(\sqrt{2}E_{a}{}^{N}\partial_{N}-\frac{1}{4}\omega_{a\bar{c}\bar{d}}
  \gamma^{\bar{c}\bar{d}}\big)\epsilon_1-(1 \leftrightarrow 2)
  \;,
 \end{split}
 \ee 
where we used that only the variation of $\Psi_a$ is non-trivial as a consequence of $\Delta_{\epsilon} E_{\bar{a}\bar{b}}=0$. 
The first term in here reads 
 \be\label{SUSysTep}
  -\frac{1}{\sqrt{2}}\big(\bar{\epsilon}_{2}\gamma_{\bar{b}}\partial_{N}\epsilon_1-
  \bar{\epsilon}_{1}\gamma_{\bar{b}}\partial_{N}\epsilon_2\big)E_{a}{}^{N} \ = \ 
   \frac{1}{\sqrt{2}}\partial_{N}\big(\bar{\epsilon}_{1}\gamma_{\bar{b}}\epsilon_2\big)E_{a}{}^{N}\;. 
 \ee 
For the second term we use as above that the $\gamma^{(3)}$ structure drops due to the antisymmetrization in $(1 \leftrightarrow 2)$. 
The remaining structure proportional to $\gamma^{(1)}$ is then automatically antisymmetric in  $(1 \leftrightarrow 2)$ and thus 
reads
 \be\label{SUsyStep}
  -\frac{1}{2}\omega_{a\bar{c}\bar{d}}\,\bar{\epsilon}_2\,\delta_{\bar{b}}{}^{[\bar{c}}\gamma^{\bar{d}]}\epsilon_1 \ = \ 
  \frac{1}{2}\omega_{a\bar{b}\bar{c}}\,\bar{\epsilon}_1\,\gamma^{\bar{c}}\epsilon_2 \;.
 \ee 
The spin connection is given by 
 \be
  \omega_{a\bar{b}\bar{c}} \ = \ -3f_{[a\bar{b}\bar{c}]} \ = \ 
  \sqrt{2}\left(E_{a}{}^{K} E_{\bar{b}}{}^{N}\partial_{K}E_{\bar{c}N}
  -E_{\bar{b}}{}^{K}\partial_{K}E_{\bar{c}}{}^{N} E_{aN}-E_{\bar{c}}{}^{K}E_{\bar{b}}{}^{N}\partial_{K} E_{aN}\right)\;.
 \ee 
Inserting this into (\ref{SUsyStep}) and combining with (\ref{SUSysTep}) we obtain in total 
 \be\label{barproj} 
  E_{\bar{b}}{}^{M}\big[ \delta_{\epsilon_1},\delta_{\epsilon_2}\big] E_{aM} \ = \ 
  E_{\bar{b}}{}^{M}\big(\xi^{N}\partial_{N}E_{aM}+\big(\partial_{M}\xi^{N}-\partial^{N}\xi_{M}\big)E_{aN}\big)\;,  \ee 
where 
 \be\label{xipara}
  \xi^{M} \ = \ -\frac{1}{\sqrt{2}}E_{\bar{a}}{}^{M}\bar{\epsilon}_1\,\gamma^{\bar{a}}\epsilon_2\;,
 \ee
is the same parameter as in (\ref{dclosure}).   

Next, we turn to the other projection, 
 \be\label{2ndproj}
 \begin{split}
  E_{b}{}^{M}\big[ \delta_{\epsilon_1},\delta_{\epsilon_2}\big] E_{aM} \ &= \ -\frac{1}{2}E_{b}{}^{M}\delta_{\epsilon_1}E_{M}{}^{\bar{c}}\,\bar{\epsilon}_{2}\gamma_{\bar{c}}
  \Psi_{a} -(1 \leftrightarrow 2)   \\
  \ &= \ \frac{1}{2}\Delta_{\epsilon_1}E_{b\bar{c}}\,\bar{\epsilon}_2\gamma^{\bar{c}}\Psi_a-(1 \leftrightarrow 2) 
  \ = \ \frac{1}{2}\big(\bar{\epsilon}_1\gamma_{\bar{c}}\Psi_{[a}\big)\big(\bar{\epsilon}_2\gamma^{\bar{c}}\Psi_{b]}\big)\;.
 \end{split}
 \ee 
The last term is antisymmetric in $a,b$ and can thus be interpreted as a field-dependent $O(1,9)_L$ gauge transformation. 
Here we would have expected also a generalized diffeomorphism with parameter (\ref{xipara}), 
but for this particular projection such a term can actually be absorbed into an $O(1,9)_L$ gauge transformation. To show
this it suffices to note that by definition (\ref{gendiff})
 \be
  E_{b}{}^{M}\delta_{\xi} E_{aM} \ = \ \xi^{N}E_{b}{}^{M}\partial_{N}E_{aM}-2\partial_{M}\xi_{N}E_{[a}{}^{M}E_{b]}{}^{N}\;,
 \ee
is antisymmetric in $a,b$. Thus, equivalently, (\ref{2ndproj}) closes into the required generalized diffeomorphisms  and into 
local  $O(1,9)_L$ transformations with parameter  
 \be\label{Lambdapara}
  \Lambda_{ab} \ = \ \frac{1}{2}\big(\bar{\epsilon}_1\gamma_{\bar{c}}\Psi_{[a}\big)\big(\bar{\epsilon}_2\gamma^{\bar{c}}\Psi_{b]}\big)
  +\xi^{N}E_{[a}{}^{M}\partial_{N}E_{b]M}+2\partial_{M}\xi_{N}E_{[a}{}^{M}E_{b]}{}^{N} \;, 
 \ee
with $\xi^{M}$ given by (\ref{xipara}).  In total, combining (\ref{barproj}) and (\ref{2ndproj}), we have verified closure,  
 \be
   \big[ \delta_{\epsilon_1},\delta_{\epsilon_2}\big] E_{aM}  \ = \ \widehat{\cal L}_{\xi}E_{aM}+\Lambda_{a}{}^{b}E_{bM}\;, 
 \ee  
with parameters given by (\ref{xipara}) and (\ref{Lambdapara}). The verification for $E_{\bar{a}M}$ is completely analogous. In particular, the 
corresponding $O(1,9)_{R}$ parameter is given by 
\be \label{Lambdabarpara}
  \Lambda_{\bar{a} \bar{b}} \ = \ \frac{1}{2}\big(\bar{\epsilon}_1\gamma_{[\bar{a}}\Psi^{c}\big)\big(\bar{\epsilon}_2\gamma_{\bar{b}]}\Psi_{c}\big)
  +\xi^{N}E_{[\bar{a}}{}^{M}\partial_{N}E_{\bar{b}]M}+2\partial_{M}\xi_{N}E_{[\bar{a}}{}^{M}E_{\bar{b}]}{}^{N} \;.
\ee
In general, the supersymmetry transformations close according to 
\be\label{gaugelag}
\big[ \delta_{\epsilon_1} , \delta_{\epsilon_2}\big]  \ = \  \delta_{\xi} + \delta_{\Lambda} + \delta_{\bar{\Lambda}} \; ,
\ee
with $\xi$ given by (\ref{xipara}), $\Lambda$ by (\ref{Lambdapara}) and $\bar{\Lambda}$ by (\ref{Lambdabarpara}).
We finally note that even though we have not employed the field equations for the above computation,  
in general the gauge algebra (\ref{gaugelag}) will only hold on-shell. In fact, without auxiliary fields 
supersymmetry transformations close on the fermions only modulo their field equations. In contrast, 
for the bosons the field equations do not enter on dimensional grounds, because they are second-order 
in derivatives.

\subsection{Reduction to standard ${\cal N}=1$ supergravity}
Let us now verify that the action (\ref{N=1DFTactionMain}) and the supersymmetry rules (\ref{DFTSUSY})
reduce to the conventional ${\cal N}=1$ supergravity in $D=10$ upon setting $\tilde{\partial}^i=0$.
As discussed above, this comparison requires a partial gauge fixing of the local $O(1,9)_L\times O(1,9)_R$ to
the diagonal subgroup. We can then write the frame field as in (\ref{coset}) in terms of $b_{ij}$ and the conventional
vielbein $e_i{}^{a}$. In the following we will show that the conventional ${\cal N}=1$ theory
is related to the action following from (\ref{N=1DFTactionMain}) by a field redefinition.

We start by recalling minimal ${\cal N}=1$, $D=10$ supergravity in the string frame.
The field content is given by
\be
\left(e_{i}{}^{a} \;, \; b_{ij}\;, \;\phi\;, \;\psi_i  \;,  \;\lambda  \right) \; ,
\ee
where the fermionic fields are the gravitino $\psi_i$ and the dilatino $\lambda$.
The action reads\footnote{This form of the
supergravity action is $\frac{1}{2}$ times the one obtained from eq.~(10) of \cite{Bergshoeff:1988nn} by performing the redefinitions 
$\phi^{-\frac{3}{2}} \rightarrow e^{- \phi} ,\,  \lambda \rightarrow \sqrt{2} \lambda, \, F_{ijk} \rightarrow \frac{1}{3 \sqrt{2} } H_{ijk},\,
B_{ij} \rightarrow \frac{1}{\sqrt{2}} b_{ij}$.}
\be \label{origaction}
\begin{split}
S  \ =  \
\int d^{10} x \,  e \, e^{-2 \phi} \Big[& \Big( R + 4 \partial^i \phi \, \partial_i \phi  - \frac{1}{12} H^{ijk} H_{ijk} \Big) \\
&- \bar{\psi}_{i} \gamma^{i j k}  D_{j} \psi_k + 2 \bar{\psi}^{i} (\partial_i \phi ) \gamma^{j} \psi_j
- 2 \bar{\lambda} \gamma^{i} D_{i} \lambda - \bar{\psi}_{i} (\slashed\partial \phi) \gamma^{i} \lambda \\
&+ \frac{1}{24} H_{ijk} \left( \bar{\psi}_{m} \gamma^{m i j k n} \psi_{n} + 6 \bar{\psi}^{i} \gamma^{j} \psi^{k} - 2 \bar{\psi}_{m} \gamma^{ijk} \gamma^{m} \lambda \right) \Big] \; ,
\end{split}
\ee
where $H_{ijk} = 3 \partial_{[i} b_{jk]}$ and $e=\det(e_{i}{}^{a})$.
Here, we denoted the covariant derivatives with respect to the standard torsion-free Levi-Civita connection 
by $D_i$ in order to distinguish them from the covariant derivatives $\nabla$ with respect to Siegel's connections.
If a non-trivial connection, say $\hat{\omega}$, is used this will be indicated explicitly as $D_{i}(\hat{\omega})$.  
We stress that the spin connection defining the Ricci scalar and thus the Einstein-Hilbert term 
is also the conventional torsion-free connection rather than the super-covariant one. 
We will not take into account terms higher order in fermions.
Up to this order, the supersymmetry transformations leaving (\ref{origaction}) invariant  read
\be\label{susyvar}
\begin{split}
\delta_{\epsilon } e_{i}{}^{a} \ &= \ \frac{1}{2}  \bar{\epsilon} \, \gamma^a \psi_i - \frac{1}{4} \bar{\epsilon} \,  \lambda \, e_{i}{}^{a} \; , \\
\delta_{\epsilon} \phi \ &= \ - \bar{\epsilon} \lambda \; , \\
\delta_{\epsilon} \psi_i \ &= \ D_{i} \epsilon  - \frac{1}{8} \gamma_i ( \slashed\partial \phi) \epsilon + \frac{1}{96} (\gamma_{i}{}^{klm} - 9 \delta_{i}{}^{k} \gamma^{lm}) H_{klm} \epsilon \; , \\
\delta_{\epsilon} \lambda \ &= \  - \frac{1}{4} (\slashed\partial \phi) \epsilon + \frac{1}{48} \gamma^{ijk} H_{ijk} \epsilon \; , \\
\delta_{\epsilon} b_{ij} \ &= \ \frac{1}{2} (\bar{\epsilon} \, \gamma_{i}  \psi_{j} - \bar{\epsilon}  \, \gamma_{j} \psi_{i}  ) - \frac{1}{2} \bar{\epsilon} \, \gamma_{ij} \lambda \; .
\end{split}
\ee

Next, we perform some field redefinitions that are necessary in order to compare with the double field theory
variables \cite{Coimbra:2011nw},
 \be\label{Psiredef}
   \Psi_i \ \equiv \ \psi_i - \frac{1}{2} \gamma_i \lambda \;, \qquad
   \rho \ \equiv  \    \gamma^i \psi_i - \lambda   \ = \ \gamma^i \Psi_i + 4 \lambda \;.
 \ee
Moreover, as usual we introduce the T-duality invariant dilaton $e^{-2d}=e\,e^{-2\phi}$.
Written in terms of these variables, the action (\ref{origaction}) reads
\be\label{finalaction}
\begin{split}
S  \ = \  \int d^{10} x \,  e^{-2 d} \Big[& \Big( R + 4 \partial^i \phi \, \partial_i \phi  - \frac{1}{12} H^{ijk} H_{ijk} \Big) - \bar{\Psi}^{j} \gamma^{i}  D_{i} \Psi_j  + 2 \bar{\Psi}^{i} D_{i} \rho  \\  &  + \bar{\rho} \gamma^i D_{i} \rho
 + \frac{1}{4} \bar{\Psi}^{i} \slashed{H} \Psi_{i} - \frac{1}{4} \bar{\rho} \slashed{H} \rho + \frac{1}{2} H_{ijk} \bar{\Psi}^{i} \gamma^{j} \Psi^{k}  + \frac{1}{4} H_{ijk} \bar{\rho} \gamma^{ij} \Psi^{k}  \Big] \; ,
\end{split}
\ee
where $\slashed{H} =  \frac{1}{3!} \gamma^{ijk} H_{ijk}$.  This is the
final form of the action that is suitable for the comparison with double field theory.
The supersymmetry variations written in terms of (\ref{Psiredef}) are
\be \label{newSUSYvar}
\begin{split}
\delta_{\epsilon} e_{i}{}^{a} \ &= \  \frac{1}{2}  \bar{\epsilon} \, \gamma^a \Psi_i \; , \\
\delta_{\epsilon} b_{ij} \ &= \  \bar{\epsilon} \, \gamma_{[i}  \Psi_{j]}  \; , \\
\delta_{\epsilon} d \ &= \  -  \frac{1}{4} \bar{\epsilon} \rho \; , \\
\delta_{\epsilon} \Psi_i \ &= \  D_{i} (\hat{\omega}) \epsilon \; , \\
\delta_{\epsilon} \rho \ &= \  \gamma^i D_{i} \epsilon - \frac{1}{24} H_{ijk} \gamma^{ijk} \epsilon  - (\slashed{\partial} \phi) \epsilon\;, \\
\end{split}
\ee
where we introduced a redefinition of the Levi-Civita spin connection $\omega^{\rm L}$,
 \be\label{omegahat}
  \hat{\omega}_{abc} \ = \ \omega^{\rm L}_{abc} - \frac{1}{2} H_{abc} \; ,
 \ee
because this is the combination that appears naturally in double field theory, see (\ref{finaloff}).

Let us now return to the double field theory action and supersymmetry transformations (\ref{N=1DFTactionMain}) and (\ref{DFTSUSY}).
We first observe that the kinetic terms in (\ref{N=1DFTactionMain}) and (\ref{finalaction}) agree, upon
converting flat into curved indices.
We will show next that the extra terms in the action (\ref{finalaction}) and the supersymmetry rules
(\ref{newSUSYvar}) as compared to double field theory are precisely reproduced
by the non-trivial connections inside the covariant derivatives in double field theory.

We start with the supersymmetry transformations. First we note that the variation of $\psi_i$ agrees with the
double field theory variation (\ref{DFTSUSY}), because (\ref{omegahat}) coincides with (\ref{finaloff}).
Next, consider the variation of the dilatino $\rho$ in (\ref{DFTSUSY}), which reads
\be\label{delrho3}
\delta_{\epsilon}\rho \ = \ \gamma^{\bar{a}} \nabla_{\bar{a}} \epsilon \ = \
\gamma^{\bar{a}} \big(E_{\bar{a}} - \frac{1}{4} \omega_{\bar{a} \bar{b} \bar{c}} \gamma^{\bar{b} \bar{c} } \big) \epsilon \; .
\ee
We can now work out the connection term in here,
\be
\omega_{\bar{a} \bar{b}\bar{c}  } \gamma^{ \bar{a} } \gamma^{\bar{b} \bar{c} }  \ = \ \omega_{\bar{a} \bar{b} \bar{c} } \big( \gamma^{ \bar{a} \bar{b} \bar{c} } - {\cal G}^{\bar{a} \bar{b} } \gamma^{ \bar{c} } + {\cal G}^{\bar{a} \bar{c} } \gamma^{\bar{b}}\big) \ = \ \omega_{[ \bar{a} \bar{b} \bar{c} ]}  \gamma^{ \bar{a} \bar{b}  \bar{c}  }  +2 \omega_{\bar{a}  \bar{b} }{}^{\bar{a} } \gamma^{\bar{b}} \; ,
\ee
where we used that $\omega$ is antisymmetric in its last two indices. 
Insertion into (\ref{delrho3}) then yields 
 \be
  \delta_{\epsilon}\rho \ = \ \big(\gamma^{\bar{a}}E_{\bar{a}}-\frac{1}{4}\omega_{[\bar{a}\bar{b}\bar{c}]}\gamma^{\bar{a}\bar{b}\bar{c}}
  -\frac{1}{2}\omega_{\bar{a}\bar{b}}{}^{\bar{a}}\gamma^{\bar{b}}\big)\epsilon\;.
 \ee
We see that only the totally antisymmetric and trace parts of the connections enter, which in turn 
are fully determined by the constraints. This observation,  
which has first been made in \cite{Coimbra:2011nw}, will be used repeatedly below.   
Inserting now (\ref{screl}) and (\ref{tracepart}) for these determined connections we
can rewrite (\ref{delrho3}) as
\be\label{rhovar}
\delta_{\epsilon} \rho \ = \  \gamma^i D_{i} \epsilon - \frac{1}{24} H_{ijk} \gamma^{ijk} \epsilon  - (\slashed{\partial} \phi) \epsilon \; ,
\ee
which agrees with the required supersymmetry variation of $\rho$ in (\ref{newSUSYvar}).
Thus, we have shown that the supersymmetry variations of the fermions in double field theory reproduce the
transformations required by ${\cal N}=1$ supergravity. For the supersymmetry variations of the bosonic fields
consistency with double field theory is manifest for the dilaton $d$, while for the metric and $b$-field a short
computation is required: variation of (\ref{coset}) yields 
 \be\label{SUSYEVAL}
  \Delta_{\epsilon}E_{a\bar{b}} \ = \ e_{\bar{b}}{}^{i}\delta_{\epsilon}e_{ia}+e_{a}{}^{i}\delta_{\epsilon}e_{i\bar{b}}
  -\frac{1}{2}e_{a}{}^{i}e_{\bar{b}}{}^{j}\delta_{\epsilon}b_{ij} \ = \ -\frac{1}{2}\bar{\epsilon}\gamma_{\bar{b}}\Psi_{a}\;.
 \ee
Due to the relative sign in the contraction of barred indices discussed after eq.~(\ref{CLiff}) we
have to identify $\gamma_{i} = -e_{i}{}^{\bar{a}}\gamma_{\bar{a}}$. Projecting (\ref{SUSYEVAL}) 
onto its antisymmetric part we then read off $\delta_{\epsilon}b_{ij}=\bar{\epsilon}\gamma_{[i}\Psi_{j]}$, 
in precise agreement with (\ref{newSUSYvar}). In addition, the symmetric projection of (\ref{SUSYEVAL})
determines the symmetric part of the supersymmetry variation $e_{b}{}^{i}\delta_{\epsilon}e_{ia}$. 
Its antisymmetric part is undetermined, as it should be, because this freedom reflects the 
diagonal local Lorentz group that is left unbroken by the gauge-fixed form (\ref{coset}). 
It is then easy to see that, up to these local Lorentz transformations,  (\ref{SUSYEVAL}) 
yields $\delta_{\epsilon}e_{i}{}^{a}$ as in (\ref{newSUSYvar}). In total, the 
supersymmetry transformations of double field theory reduce precisely to (\ref{newSUSYvar}).

We turn now to the action. Similarly to the discussion of the supersymmetry transformations it is easy to see
that all connections are determined and that writing them out in terms of the Levi-Civita connection reproduces
the $H$-dependent terms in (\ref{finalaction}).

Let us start with the covariant derivative $\nabla_{\bar{b}}$ in the first fermionic term in (\ref{N=1DFTactionMain}), which acts on
$\Psi_{a}$ as an $O(1,9)_{R}$ spinor and as an $O(1,9)_{L}$ vector, i.e.,
 \be\label{STEPPPP}
  -\bar{\Psi}^{a}\gamma^{\bar{b}}\nabla_{\bar{b}}\Psi_{a} \ = \ -\bar{\Psi}^{a}\gamma^{\bar{b}}\big(E_{\bar{b}}\Psi_{a}-\frac{1}{4}\omega_{\bar{b}\bar{c}\bar{d}}\gamma^{\bar{c}\bar{d}}\Psi_{a}
  +\omega_{\bar{b}a}{}^{c}\Psi_{c}\big)\;.
 \ee
As in (\ref{rhovar}), the terms combine into $-\bar{\Psi}^{j}\gamma^{i} D_{i}\Psi_{j}$ and $\tfrac{1}{4}\bar{\Psi}^{i}\slashed{H}\Psi_{i}$, while
a $d$-dependent term drops out as a consequence of $\Psi^{j}\gamma^{i}\Psi_{j}=0$. The last term in (\ref{STEPPPP}) gives
in addition to the standard spin connection an extra contribution,  
\be\label{RedStep1}
-\bar{\Psi}^a \gamma^{\bar{b}} \omega_{\bar{b} a }{}^{c} \, \Psi_{c} \ = \
-\bar{\Psi}^a \gamma^{\bar{b}}\Big( \omega^{\rm L}_{\bar{b} a }{}^{c} + \frac{1}{2 }  H_{\bar{b} a}{}^{c} \Big)  \Psi_{c}
\ = \  -\bar{\Psi}^a \gamma^{\bar{b}}\omega^{\rm L}_{\bar{b} a }{}^{c} \,  \Psi_{c}  + \frac{1}{2 }  H_{ a \bar{b} c} \bar{\Psi}^a \gamma^{\bar{b}}  \Psi^{c} \; , 
\ee
reproducing the term $\tfrac{1}{2}H_{ijk}\bar{\Psi}^i\gamma^{j}\Psi^{k}$ in (\ref{finalaction}).

Next, we consider the kinetic term of $\rho$ which as in (\ref{rhovar}) reduces to
 \be\label{RedStep2}
  \bar{\rho}\gamma^{\bar{a}}\nabla_{\bar{a}}\rho \ = \ \bar{\rho}\gamma^{i}D_{i}\rho-\frac{1}{24}\bar{\rho}H_{ijk}\gamma^{ijk}\rho\;.
 \ee
Finally, the last structure in (\ref{N=1DFTactionMain}) yields
 \be\label{RedStep3}
  2\bar{\Psi}^{a}\nabla_{a}\rho \ = \ 2\bar{\Psi}^{a}\big(E_{a}\rho-\frac{1}{4}\omega_{a\bar{b}\bar{c}}\gamma^{\bar{b}\bar{c}}\rho\big)
  \ = \ 2\bar{\Psi}^{i}D_{i}\rho+\frac{1}{4}H_{ijk}\bar{\rho}\gamma^{ij}\Psi^{k}\;.
 \ee
Collecting the term $\tfrac{1}{4}\bar{\Psi}^{i}\slashed{H}\Psi_{i}$ originating from (\ref{STEPPPP}) together with 
(\ref{RedStep1}), (\ref{RedStep2}) and (\ref{RedStep3}) we infer that the double field theory
action reproduces (\ref{finalaction}).
Summarizing, we have shown that the ${\cal N}=1$ supersymmetric double field theory reduces for $\tilde{\partial}^i=0$
to minimal ${\cal N}=1$ supergravity in $D=10$.

\section{Heterotic Supersymmetric Double Field Theory}
In this section we extend the above construction to the coupling of an arbitrary number $n$ of
abelian vector multiplets. For $n=16$ this completes the construction of \cite{Hohm:2011ex}
by the fermionic or NS-R sector of heterotic superstring theory truncated to the
Cartan subalgebra of $E_8\times E_8$ or $SO(32)$.
We first review the extension of the frame formalism, in which the
tangent space group is extended to $O(1,9+n)\times O(1,9)$. Then we show that the same
${\cal N}=1$ double field theory action (\ref{N=1DFTaction}), but interpreted with respect to
the enlarged frame and spinor fields, reduces to ${\cal N}=1$
supergravity coupled to $n$ vector multiplets upon setting the extra derivatives to zero.

\subsection{${\cal N}=1$ double field theory with local $O(1,9+n)\times O(1,9)$ symmetry}
Let us begin by reviewing the double field theory formulation in presence of $n$
abelian gauge vectors $A_{i}{}^{\alpha}$ \cite{Hohm:2011ex}. The generalized metric is extended to
an $O(10+n,10)$ group element, naturally encoding these additional fields.
Correspondingly, there are $20+n$ coordinates,
 \be
  X^{M} \ = \ (\tilde{x}_{i}\,,\; y^{\alpha}\,, \;x^{i})\;, \qquad 
  \partial_{M} \ = \ (\tilde{\partial}^{i}\,,\; \partial_{\alpha}\,, \;\partial_{i})\;, 
 \ee
transforming as an $O(10+n,10)$ vector, with indices that are raised and lowered with
 \be\label{bigeta}
  \eta_{MN} \ = \   \begin{pmatrix} 0 & 0 & {\bf 1}_{10} \\
  0 & {\bf 1}_{n} & 0 \\ {\bf 1}_{10} & 0 &  0
 \end{pmatrix}\;.
\ee
We still impose the constraint $\eta^{MN}\partial_{M}\partial_{N}=0$, 
using the $O(10+n,10)$ invariant metric (\ref{bigeta}). It implies 
that one can always rotate into a frame in which $\tilde{\partial}^i=\partial_{\alpha}=0$.  

Next, we can introduce an enlarged frame field as in (\ref{framedef}),
but now with indices $a,b,\ldots$ taking $10+n$ values and with the
upper-left block of $\hat{\eta}_{AB}$ being
\be
 \eta_{ab}  \  = \ \begin{pmatrix} \eta_{\underline{a}\underline{b}} & 0  \\
0 & \delta_{\underline{\alpha}\underline{\beta}} \\
 \end{pmatrix}\;.
\ee
Here and in the following we split flat indices as
 \be
  A \ = \ (a\,,\;\bar{a}) \ = \ (\underline{a}\,,\;\underline{\alpha}\,,\; \bar{a})\;, \qquad \underline{a}=0,\ldots,9\;, \quad \underline{\alpha}=1,\ldots,n\;.
 \ee
The frame field is constrained by requiring that the tangent space metric ${\cal G}_{AB}$ still satisfies
(\ref{Gconstr}), which reads explicitly
  \be\label{GconstrH}
  {\cal G}_{a\bar{b}} \ = \ 0\;, \qquad
  {\cal G}_{\underline{a}\underline{b}} \ = \ \eta_{\underline{a}\underline{b}}\;, \qquad
  {\cal G}_{\bar{a}\bar{b}} \ = \ -\eta_{\bar{a}\bar{b}}\;, \qquad
  {\cal G}_{\underline{\alpha}\underline{\beta}} \ = \ \delta_{\underline{\alpha}\underline{\beta}}\;.
 \ee
We can then choose a gauge and parametrize the frame field as follows
\be \label{hetparam}
E_{A}{}^{M} \ = \
\begin{pmatrix} E_{\underline{a}i} & E_{\underline{a}}{}^{\beta} &   E_{\underline{a}}{}^{i}  \\
E_{\underline{\alpha}i} & E_{\underline{\alpha}}{}^{\beta} & E_{\underline{\alpha}}{}^{i}  \\
E_{\bar{a}i} & E_{\bar{a}}{}^{\beta} & E_{\bar{a}}{}^{i} \end{pmatrix} \ = \
\frac{1}{\sqrt{2}} \begin{pmatrix} e_{i\underline{a}} - e_{\underline{a}}{}^{k} c_{ki} & - e_{\underline{a}}{}^{k} A_{k}{}^{\beta}  & e_{\underline{a}}{}^{i}\\
  \sqrt{2} A_{i \underline{\alpha}} & \sqrt{2} \delta_{\underline{\alpha}}{}^{\beta} &0 \\
 - e_{i\bar{a}} - e_{\bar{a}}{}^{k} c_{ki} & - e_{\bar{a}}{}^{k} A_{k}{}^{\beta} &e_{\bar{a}}{}^{i}
 \end{pmatrix}
 \; , 
\ee
where we defined $c_{ij} = b_{ij} + \frac{1}{2} A_{i}{}^{\alpha} A_{j}{}_{\alpha}$, 
and we freely raise and lower gauge group indices with the Kronecker delta $\delta_{\alpha\beta}$.

All results of the frame formalism reviewed in sec.~\ref{framesec} extend directly to the present generalization.
In particular, all statements about determined connection components can be readily applied.
Moreover, the supersymmetric extension (\ref{N=1DFTactionMain}) is well-defined for these extended 
fields in that the gamma matrices $\gamma^{\bar{a}}$ and all spinor indices are still to be interpreted  
with respect to $O(1,9)$. The check of supersymmetric invariance and closure of the supersymmetry transformations 
immediately generalizes to the present case, as it is never used whether $a$ takes $10$ or $10+n$ values. 
Assuming the parametrization (\ref{hetparam}) and setting $\tilde{\partial}^i=\partial_{\alpha}=0$ we compute the
following connection components:
\be \label{hetspincon}
\begin{split}
\omega_{\underline{a} \bar{b} \bar{c}}  \ = \ & - \big(\omega^{\rm L}_{\underline{a} \bar{b} \bar{c}}(e) - \frac{1}{2} \hat{H}_{\underline{a} \bar{b} \bar{c}}\big) \;, \qquad
\omega_{\bar{a} \underline{b}\underline{c}}  \ = \  \omega^{\rm L}_{\bar{a} \underline{b}\underline{c} }(e) + \frac{1}{2} \hat{H}_{\bar{a} \underline{b}\underline{c}} \; ,
\\
\omega_{[\bar{a} \bar{b} \bar{c}]} \ = \ &- \big(\omega^{\rm L}_{[\bar{a} \bar{b} \bar{c}]}(e) - \frac{1}{6} \hat{H}_{\bar{a} \bar{b} \bar{c} }\big) \; , \\
\omega^{\underline{\alpha}}{}_{\bar{b} \bar{c}} \ = \ & \frac{1}{\sqrt{2}} F_{\bar{b} \bar{c}}{}^{\underline{\alpha}} \; , \qquad
\omega_{\bar{b} \underline{a}}{}^{\underline{\alpha}} \ = \ - \omega_{\bar{b}}{}^{\underline{\alpha}}{}_{\underline{a}}
\ = \  \frac{1}{\sqrt{2}} F_{\bar{b} \underline{a}}{}^{\underline{\alpha}} \; ,
\end{split}
\ee
where
 \be\label{hatstrength}
 \begin{split}
  F_{ab}{}^{\alpha} \ &= \ e_{a}{}^{i}e_{b}{}^{j}\big(\partial_i A_{j}{}^{\alpha}-\partial_{j}A_{i}{}^{\alpha}\big)\;, \\
  \hat{H}_{abc} \ &= \  3e_{a}{}^{i}e_{b}{}^{j}e_{c}{}^{k}\big(\partial_{[i}b_{jk]}-A_{[i}{}^{\alpha}\partial_{j}A_{k]\alpha}\big)\;.
 \end{split}
 \ee
Thus, we obtained the abelian field strength of the gauge fields $A_{i}{}^{\alpha}$ and  
the required Chern-Simons modification of the field strength $H$.

\subsection{Reduction to ${\cal N}=1$ Supergravity with $n$ vector multiplets}
We will now show that the ${\cal N}=1$ double field theory action with tangent space symmetry  $O(1,9+n)\times O(1,9)$
reproduces standard ${\cal N}=1$ supergravity with $n$ abelian vector multiplets upon setting $\tilde{\partial}^i=\partial_{\alpha}=0$.
Let us first recall ${\cal N}=1$ supergravity coupled to
$n$ vector multiplets
\be
\big(A_{i}{}^{\alpha}\,,\; \chi^{\alpha} \big) \; , \qquad \alpha \ = \ 1, \ldots, n\;.
\ee
The action is given by
\be \label{hetstandard}
\begin{split}
S  \ = \  \int d^{10} x \,  e \, &e^{-2 \phi} \Big[ \Big( R + 4 \partial^i \phi \, \partial_i \phi  - \frac{1}{12} \hat{H}^{ijk} \hat{H}_{ijk} - \frac{1}{4}  
F_{\alpha ij} \,F^{\alpha  ij} \Big) \\
 &- \bar{\psi}_{i} \gamma^{i j k}  D_{j} \psi_k
 - 2 \bar{\lambda} \gamma^{i} D_{i} \lambda     -  \frac{1}{2} \bar{\chi}^{\alpha} \slashed{D} \chi_{\alpha}   \\
&+ 2 \bar{\psi}^{i} (\partial_i \phi ) \gamma^{j} \psi_j - \bar{\psi}_{i} (\slashed\partial \phi) \gamma^{i} \lambda
- \frac{1}{4} \bar{\chi}_{\alpha} \gamma^{i} \gamma^{jk} F_{jk}{}^{\alpha} \big(\psi_i + \frac{1}{6} \gamma_{i} \lambda\big)\\
&+ \frac{1}{24} \hat{H}_{ijk} \Big( \bar{\psi}_{m} \gamma^{m i j k n} \psi_{n} + 6 \bar{\psi}^{i} \gamma^{j} \psi^{k}
- 2 \bar{\psi}_{m} \gamma^{ijk} \gamma^{m} \lambda +  \frac{1}{2} \bar{\chi}^{\alpha} \gamma^{ijk} \chi_{\alpha} \Big) \Big]  \;,
\end{split}
\ee
where $\hat{H}_{ijk}$ is the $H$-field strength modified by the Chern-Simons 3-form, as in (\ref{hatstrength}).
This action is invariant under the supersymmetry transformations:
\be \label{hetsusytr}
\begin{split}
\delta_{\epsilon } e_{i}{}^{a} \ = \ & \frac{1}{2}  \bar{\epsilon} \, \gamma^a \psi_i - \frac{1}{4} \bar{\epsilon} \,  \lambda \, e_{i}{}^{a} \; , \\
\delta_{\epsilon} \phi \ = \ & - \bar{\epsilon} \lambda \quad , \quad  \delta_{\epsilon} A_{i}{}^{\alpha} \ = \  \frac{1}{2} \bar{\epsilon} \, \gamma_{i} \chi^{\alpha} \quad , \quad \delta_{\epsilon} \chi^{\alpha} \ = \  - \frac{1}{4} \gamma^{ij} F_{ij}{}^{\alpha} \epsilon  \\
\delta_{\epsilon} \psi_i \ = \ & D_{i} \epsilon  - \frac{1}{8} \gamma_i ( \slashed\partial \phi) \epsilon + \frac{1}{96} (\gamma_{i}{}^{klm} - 9 \delta_{i}{}^{k} \gamma^{lm}) \hat{H}_{klm} \epsilon \; , \\
\delta_{\epsilon} \lambda \ = \ & - \frac{1}{4} (\slashed\partial \phi) \epsilon + \frac{1}{48} \gamma^{ijk} \hat{H}_{ijk} \epsilon \; , \\
\delta_{\epsilon} b_{ij} \ = \ & \frac{1}{2} (\bar{\epsilon} \, \gamma_{i}  \psi_{j} - \bar{\epsilon}  \, \gamma_{j} \psi_{i}  ) - \frac{1}{2} \bar{\epsilon} \, \gamma_{ij} \lambda   +  \frac{1}{2} \bar{\epsilon} \gamma_{[i} \chi^{\alpha} A_{j]\alpha} \; .
\end{split}
\ee
Next, we perform the same field redefinition (\ref{Psiredef}) as for the minimal theory.
We obtain for the action
\be\label{hetIIaction}
\begin{split}
S_{F} &\ =  \  \int d^{10} x \,  e^{-2 d} \left[  - \bar{\Psi}^{j} \gamma^{i}  D_{i} \Psi_j  + 2 \bar{\Psi}^{i} D_{i} \rho   + \bar{\rho} \gamma^i D_{i} \rho  -  \frac{1}{2} \bar{\chi}^{\alpha} \gamma^{i} D_{i}  \chi_{\alpha} - \frac{1}{4}   \bar{\chi}^{\alpha}  \gamma^{jk} F_{jk\alpha} \rho
 \right. \\  &   \left.  -   \bar{\chi}^{\alpha} \gamma^{k} F_{i k\alpha} \Psi^{i}  + \frac{1}{4} \bar{\Psi}^{i} \hat{\slashed{H}} \Psi_{i}  - \frac{1}{4} \bar{\rho} \hat{\slashed{H}} \rho+ \frac{1}{2} \hat{H}_{ijk} \bar{\Psi}^{i} \gamma^{j} \Psi^{k}  + \frac{1}{4} \hat{H}_{ijk} \bar{\rho} \gamma^{ij} \Psi^{k} + \frac{1}{8}  \bar{\chi}^{\alpha} \hat{\slashed{H}} \chi_{\alpha}  \right] \; ,
\end{split}
\ee
and the supersymmetry transformations are given by
\be\label{newsusyhet}
\begin{split}
\delta_{\epsilon} e_{i}{}^{a} \ = \ & \frac{1}{2}  \bar{\epsilon} \, \gamma^a \Psi_i \;, \qquad \delta_{\epsilon} \Psi_i \ = \  D_{i} (\hat{\omega}) \epsilon  \; , \\
\delta_{\epsilon} b_{ij} \ = \ & \bar{\epsilon} \, \gamma_{[i}  \Psi_{j]}  +  \frac{1}{2} \bar{\epsilon} \gamma_{[i} \chi A_{j]} \; ,  \\
\delta_{\epsilon} d \ = \ & -  \frac{1}{4} \bar{\epsilon} \rho \;, \qquad
\delta_{\epsilon} \rho \ = \  \gamma^i D_{i} \epsilon - \frac{1}{24} \hat{H}_{ijk} \gamma^{ijk} \epsilon  - (\slashed{\partial} \phi) \epsilon\;,  \\
\delta_{\epsilon} A_{i}{}^{\alpha} \ = \ & \frac{1}{2} \bar{\epsilon} \gamma_{i} \chi^{\alpha} \quad , \quad \delta_{\epsilon} \chi^{\alpha} \ = \ - \frac{1}{4} \gamma^{ij} F_{ij}{}^{\alpha} \epsilon \; .
\end{split}
\ee

Let us now verify that the above action and supersymmetry rules are reproduced by supersymmetric double field theory
for $\tilde{\partial}^i=\partial_{\alpha}=0$.
Here, our discussion will be a little briefer than above because it suffices 
to focus on the new structures involving the gauge vectors and gauginos. 
It turns out that the comparison requires the identification
\be
\Psi_{a} \ = \ \big(\Psi_{\underline{a}}\,,\;\Psi_{\underline{\alpha}}\big) \ \equiv \
\big(e_{\underline{a}}{}^{i} \Psi_{i}\,,\;\tfrac{1}{\sqrt{2}} \chi_{\underline{\alpha}}\big)\;,
\ee
i.e., the gauginos are naturally identified with the additional components of the `gravitino'.

We start with the supersymmetry transformations. The gaugino variation $\delta_{\epsilon} \chi^{\alpha}$
can be obtained by considering 
 \be \label{hetsusy}
 \begin{split}
 \delta_{\epsilon} {\Psi}_{\underline{\alpha}} \ = \ \frac{1}{\sqrt{2}} \delta_{\epsilon} \chi_{\underline{\alpha}}
 \ = \  \nabla_{\underline{\alpha}} \epsilon
  \ = \ \left( \sqrt{2}E_{\underline{\alpha}}{}^{i} \partial_{i} \epsilon - \frac{1}{4} \omega_{\underline{\alpha} \bar{b} \bar{c}} \gamma^{\bar{b} \bar{c}} \epsilon \right)
  \  = \ - \frac{1}{4 \sqrt{2}} F_{\bar{b} \bar{c}\,\underline{\alpha}}\, \gamma^{\bar{b} \bar{c}} \epsilon \; ,
 \end{split}
 \ee
 where we used (\ref{hetspincon}) and  $E_{\underline{\alpha}}{}^{i} = 0$ for the gauge choice (\ref{hetparam}). We read off
 \be
  \delta_{\epsilon} \chi^{\underline{\alpha}} \ = \  - \frac{1}{4} F_{\bar{b} \bar{c} }{}^{\underline{\alpha}}\, \gamma^{\bar{b} \bar{c}} \epsilon \; .
 \ee
Comparison with (\ref{newsusyhet}) shows that we obtained the expected supersymmetry variation.
For the supersymmetry variations of the vielbein $e_{i}{}^{a}$, the $b$-field and the gauge vectors we compute 
as in (\ref{SUSYEVAL}) the variation of the gauge-fixed frame field (\ref{hetparam}) 
\be
\Delta_{\epsilon} E_{ \underline{a} \bar{b}} \ = \ e_{\bar{b}}{}^{i} \delta_{\epsilon} e_{i \underline{a}} + e_{\underline{a}}{}^{i} \delta_{\epsilon} e_{i \bar{b}} - \frac{1}{2} e_{\underline{a}}{}^{i} e_{\bar{b}}{}^{j} \delta_{\epsilon} b_{ij} - \frac{1}{2} e_{\underline{a}}{}^{i} e_{\bar{b}}{}^{j} A_{[i}{}^{\underline{\alpha}} \,\delta_{\epsilon} A_{j] \underline{\alpha}} \ = \ - \frac{1}{2} \bar{\epsilon} \gamma_{\bar{b}} \Psi_{\underline{a}} \; , 
\ee
and 
\be
\Delta_{\epsilon} E_{\underline{\alpha} \bar{b}} \ = \ \frac{\sqrt{2}}{2} e_{\bar{b}}{}^{i} \delta_{\epsilon} A_{i \underline{\alpha}} \ = \ -\frac{1}{2 \sqrt{2}} \bar{\epsilon} \gamma_{\bar{b}} \chi_{\underline{\alpha}} \; .
\ee
Combining these two gives the required supersymmetry transformations (\ref{newsusyhet}).

Let us now turn to the action and show that it produces the required $\chi$-dependent terms. For the first fermionic term in (\ref{N=1DFTaction})
we obtain
 \be
 \begin{split}
 - \bar{\Psi}^{a}\gamma^{\bar{b}}\nabla_{\bar{b}}\Psi_{a} \Big|_{\chi} \ = \  &
  -  \frac{1}{2} \bar{\chi}^{\underline{\alpha}} \gamma^{\bar{b}}D_{\bar{b}}   \chi_{\underline{\alpha}}
  +\frac{1}{8}\bar{\chi}^{\underline{\alpha}}\slashed{\hat{H}}\chi_{\underline{\alpha}}
 - \bar{\Psi}^{\underline{a}} \gamma^{\bar{b}} \omega_{\bar{b} \underline{a}}{}^{\underline{\alpha}}  \Psi_{\underline{\alpha}}  
 -  \bar{\Psi}^{\underline{\alpha}} \gamma^{\bar{b}} \omega_{\bar{b} \underline{\alpha}}{}^{\underline{a}}
 \Psi_{\underline{a}} \\
  = \ & -  \frac{1}{2} \bar{\chi}^{\underline{\alpha}} \gamma^{\bar{b}}D_{\bar{b}}   \chi_{\underline{\alpha}}
  +\frac{1}{8}\bar{\chi}^{\underline{\alpha}}\slashed{\hat{H}}\chi_{\underline{\alpha}}
 - \bar{\chi}_{\underline{\alpha}} \gamma^{\bar{b}} F_{\underline{a} \bar{b}}{}^{\underline{\alpha}}\, \Psi^{\underline{a}} \;,
 \end{split}
 \ee
where we used in the first line that the last two terms are equal.
 The second fermionic term in the action (\ref{N=1DFTaction}) does not give any $\chi$-dependent contribution. The third term reads
 \be
  2 \bar{\Psi}^{a}\nabla_{a}\rho \Big|_{\chi} \ = \ 2  \bar{\Psi}^{\underline{\alpha}} \nabla_{\underline{\alpha}} \rho \ = \
  \frac{2}{\sqrt{2}}  \bar{\chi}^{\underline{\alpha}} \Big( - \frac{1}{4} \omega_{\underline{\alpha} \bar{b} \bar{c}} \gamma^{\bar{b} \bar{c}}   \Big) \rho \ = \ - \frac{1}{4 } \bar{\chi}_{\underline{\alpha}} \gamma^{\bar{b} \bar{c}} F_{\bar{b} \bar{c}}{}^{\underline{\alpha}}\, \rho \;,
 \ee
reproducing the required coupling in (\ref{hetIIaction}).
Thus, we have shown that all new $\chi$- and $F$-dependent terms due to the coupling of vector multiplets are precisely
reproduced by the extended connections of the $O(1,9+n)\times O(1,9)$ tangent space symmetry.

\section{Conclusions}
In this paper we have constructed the ${\cal N}=1$ supersymmetric extension of double field theory 
for $D=10$. This theory features two copies of the local Lorentz group as tangent space symmetries, 
under which the fermions naturally transform. Interestingly, the generalization to the 
coupling of $n$ abelian vector multiplets amounts only to the extension of the T-duality 
group to $O(10+n,10)$ and, correspondingly, to the extension of the tangent space group 
to $O(1,9+n)\times O(1,9)$. The `gravitino' $\Psi_a$ thereby receives $n$ additional components
that can be identified with the gauginos. Apart from exhibiting a further `unification' 
of the massless sector of heterotic superstring theory, this formulation provides a 
significant technical simplification of the effective action, as should be 
apparent by comparing (\ref{hetstandard}) with (\ref{N=1DFTaction}).
Moreover, the proof of supersymmetric invariance (up to the higher order fermi terms)
is much simpler than in the standard formulation, being essentially reduced to a
two-line calculation in (\ref{fermvar}). 

On a technical level it is interesting to note that the connections emerging naturally 
in double field theory, $\omega^{\pm}=\omega^{\rm L}\pm\tfrac{1}{2}H$, have appeared in different 
contexts in string theory. For instance, they turn out to be very useful for constructing 
supersymmetric higher-derivative invariants \cite{Bergshoeff1989}, and it would be 
interesting to understand the significance of this relation. 

This work can be extended into many directions. First, the generalization to non-abelian 
vector multiplets is necessary in order to describe the full massless sector of the 
heterotic superstring. For the bosonic sector we described in \cite{Hohm:2011ex} also
the non-abelian generalization, but the formalism and physical interpretation is 
different. We hope to come back to this problem. 

Next, one should construct the ${\cal N}=2$ supersymmetric extension of the type II 
double field theory constructed in \cite{Hohm:2011zr}. The recent work 
\cite{Coimbra:2011nw} completes the corresponding construction in generalized geometry,
but there are a few subtleties in double field theory that we hope to address and 
resolve in the near future. 

Finally, the recent results \cite{Berman:2010is} on similar constructions for M-theory or 
11-dimensional supergravity suggest that an analogous supersymmetric extension 
is possible there. Here we note that in \cite{Hillmann:2009ci} the supersymmetry variations 
of 11-dimensional supergravity (in a certain truncation to $D=7$) have already been 
written in an $E_{7(7)}$ and $SU(8)$ covariant way, and it would be nice to show that the corresponding 
action is supersymmetric modulo a covariant constraint.

\section*{Acknowledgments}
We would like to thank Barton Zwiebach for many helpful discussions.

This work is supported by the U.S. Department of Energy (DoE) under the cooperative
research agreement DE-FG02-05ER41360, the  
DFG Transregional Collaborative Research Centre TRR 33
and the DFG cluster of excellence "Origin and Structure of the Universe". 
The work of SK is supported in part by a Samsung Scholarship.

 \appendix

\section{Identities for the curvature tensors}
\setcounter{equation}{0}
In this appendix we present some details of the derivation of the identities (\ref{CurvIdent}).
We start with the second one, involving the Ricci tensor, and compute
 \be
\begin{split}
[\gamma^{\bar{a}} \nabla_{\bar{a}} , &\nabla_{b} ] \epsilon
 \ = \  \gamma^{\bar{a}} \nabla_{\bar{a}} \nabla_{b} \epsilon - \nabla_{b}\big( \gamma^{\bar{a}} \nabla_{\bar{a}}  \epsilon\big)  \\
 \ =  \   &\Big( \gamma^{\bar{a}} E_{\bar{a}} - \frac{1}{4} \omega_{\bar{a} \bar{e} \bar{f}} \gamma^{\bar{a} \bar{e} \bar{f} } - \frac{1}{2} \omega_{\bar{a} \bar{e} }{}^{\bar{a}} \gamma^{\bar{e}}   \Big) \Big( E_{b} - \frac{1}{4} \omega_{  b  \bar{c} \bar{d} } \gamma^{ \bar{c} \bar{d} }  \Big) \epsilon
   + \gamma^{\bar{a}} \omega_{\bar{a}    b    } {}^{  f  }  \Big( E_{f } - \frac{1}{4} \omega_{ f \bar{c} \bar{d} } \gamma^{ \bar{c} \bar{d} }  \Big) \epsilon \\
   & -  \Big( E_{b} - \frac{1}{4} \omega_{  b  \bar{c} \bar{d} } \gamma^{ \bar{c} \bar{d} }  \Big)  \Big( \gamma^{\bar{a}} E_{\bar{a}} - \frac{1}{4} \omega_{\bar{a} \bar{e} \bar{f}} \gamma^{\bar{a} \bar{e} \bar{f} } - \frac{1}{2} \omega_{\bar{a} \bar{e} }{}^{\bar{a}} \gamma^{\bar{e}}   \Big) \epsilon \; .
\end{split}
\ee
Our strategy is now to work out the various powers $\gamma^{(p)}$ of gamma matrices separately and to show that all except
$\gamma^{(1)}$ cancel. The non-vanishing contribution will then be shown to be related to the Ricci tensor.
To this end we use the following identities for the product of (antisymmetrized) gamma matrices
\bea
\gamma^{\bar{a}} \gamma_{\bar{b}} &=& \gamma^{\bar{a}}{}_{\bar{b}} - \delta^{\bar{a}}{}_{\bar{b}} \; , \\
\gamma^{\bar{a} \bar{b}} \gamma_{\bar{c}} &=&  \gamma^{\bar{a} \bar{b}}{}_{\bar{c}} + 2 \delta^{[ \bar{a}}{}_{\bar{c}}\, \gamma^{\bar{b}]} \; , \\ 
\gamma^{\bar{a} \bar{b} \bar{c}} \gamma_{\bar{d}} &=&  \gamma^{\bar{a} \bar{b} \bar{c}}{}_{\bar{d}} - 3\delta^{[\bar{a}}{}_{\bar{d}}\,   \gamma^{\bar{b} \bar{c}]} \; , \\
\gamma^{\bar{a} \bar{b} } \gamma_{\bar{c} \bar{d}} & = & \gamma^{\bar{a} \bar{b} }{}_{\bar{c} \bar{d}} + 4\delta^{[\bar{a}}{}_{[\bar{c}}\, \gamma^{\bar{b}]}{}_{\bar{d}]} 
- 2 \delta^{[\bar{a}}{}_{[\bar{c}}\, \delta^{\bar{b}]}{}_{\bar{d}]} \; , \\
\gamma^{\bar{a} \bar{b} \bar{c}} \gamma_{\bar{d} \bar{e}} & = &  \gamma^{\bar{a} \bar{b} \bar{c}}{}_{\bar{d} \bar{e}} - 6 \delta^{[\bar{a}}{}_{[\bar{d}}\,   \gamma^{\bar{b} \bar{c}]}{}_{\bar{e}]} 
- 6  \delta^{[\bar{a}}{}_{[\bar{d}}\,   \delta^{\bar{b}}{}_{\bar{e}]}\,  \gamma^{\bar{c}]} \; ,   \\
\gamma^{\bar{a} \bar{b} \bar{c}} \gamma_{\bar{d} \bar{e} \bar{f}} & = & \gamma^{\bar{a} \bar{b} \bar{c}}{}_{\bar{d} \bar{e} \bar{f}} - 9  \delta^{[\bar{a}}{}_{[\bar{d}}\,  \gamma^{\bar{b} \bar{c}] }{}_{ \bar{e} \bar{f}]} - 18 \delta^{[\bar{a}}{}_{[\bar{d}}\, \delta^{\bar{b}}{}_{\bar{e}}\,  \gamma^{\bar{c} ] }{}_{  \bar{f}]} + 6  \delta^{[\bar{a}}{}_{[\bar{d}}\, \delta^{\bar{b}}{}_{\bar{e}}\,  \delta^{\bar{c} ] }{}_{  \bar{f}]} \; ,
 \eea
where we recall that indices are raised and lowered with ${\cal G}_{\bar{a}\bar{b}}=-\eta_{\bar{a}\bar{b}}$. 

Let us now start the computation. 
First, the $\gamma^{(5)}$ terms cancel:
\be
\frac{1}{16}  \omega_{\bar{a} \bar{e} \bar{f}}  \omega_{  b  \bar{c} \bar{d} } \gamma^{\bar{a} \bar{e} \bar{f} \bar{c} \bar{d}}  - \frac{1}{16}  \omega_{\bar{a} \bar{e} \bar{f}}  \omega_{  b  \bar{c} \bar{d} } \gamma^{  \bar{c} \bar{d} \bar{a} \bar{e} \bar{f} } \ = \ 0 \; .
\ee
Second, it is easy to see by inspection that there are no $\gamma^{(4)}$ terms.
Next, collecting terms with $\gamma^{(3)}$  we find
\be
\begin{split}
\Big[ &- \frac{1}{4} E_{\bar{a}} \omega_{b \bar{c} \bar{d}}  + \frac{1}{4} E_{b} \omega_{[\bar{a} \bar{c} \bar{d}  ]} + \frac{3}{4} \omega_{[\bar{e} \bar{c} \bar{d}]}\, \omega_{b \bar{a}}{}^{\bar{e}}   -  \frac{1}{4} \omega_{e \bar{c} \bar{d} }\, \omega_{\bar{a} b}{}^{e} \Big] \gamma^{\bar{a} \bar{c} \bar{d}} \epsilon \\
& \ = \   \frac{1}{4} \left[ E_{\bar{a} } \Omega_{b \bar{c} \bar{d}}  - E_{b}  \Omega_{[\bar{a} \bar{b} \bar{d}]}  - \Omega_{[\bar{e} \bar{c} \bar{d}]}\, \Omega_{\bar{a} b }{}^{\bar{e}}  - \Omega_{e \bar{c} \bar{d} }\,  \Omega_{\bar{a} b }{}^{e}  \right] \gamma^{\bar{a} \bar{c} \bar{d}} \epsilon \; ,
\end{split}
\ee
where we inserted in the second line the solutions for the connections.
Inserting now the explicit expressions for $\Omega$ it is a straightforward though somewhat lengthy
calculation to verify that this vanishes. It is again easy to see that there are
no $\gamma^{(2)}$ terms. So we finally have to work out the terms proportional to $\gamma^{(1)}$, for which 
we find
 \be
\begin{split}
  \gamma^{\bar{a}} \big[ & \left( (E_{\bar{a}} E_{b}{}^{M})E_{M}{}^{C} -  (E_{ b } E_{ \bar{a} }{}^{M}) E_{M}{}^{C}  \right) E_{C} + \omega_{\bar{a} b}{}^{c} E_c  -  \omega_{b \bar{a}}{}^{\bar{c}} E_{\bar{c}}  \big]  \epsilon \\
 & - \frac{1}{2} \gamma^{\bar{a}} \left[ E_{\bar{c}} \omega_{b \bar{a}}{}^{\bar{c}} - E_{b} \omega_{\bar{c} \bar{a} }{}^{\bar{c}} + \omega_{d \bar{a} }{}^{\bar{c}}\, \omega_{\bar{c} b} {}^{d} - \omega_{b \bar{a}}{}^{ \bar{d} }\, \omega_{\bar{c} \bar{d}}{}^{\bar{c}} \right] \epsilon \; .
\end{split}
\ee
The terms in the first line vanish as a consequence of the torsion constraint
(\ref{NewTorsion}):  Using $f_{ABC} \equiv (E_A E_{B}{}^{M}) E_{CM}$, the torsion constraint reads
 \be
 (f_{\bar{a} b}{}^{C} - f_{b \bar{a}}{}^{C}) E_{C} + \omega_{\bar{a} b}{}^{c} E_c  -  \omega_{b \bar{a}}{}^{\bar{c}} E_{\bar{c}}
 \ = \ (\Omega_{\bar{a} b }{}^{C} +2 \omega_{ [\bar{a} b] }{}^{C}) E_{C} \ = \ 0 \; ,
 \ee
 where we used the strong constraint (\ref{strongconstr}). Thus the final result is
 \be
 \begin{split}
 [\gamma^{\bar{a}} \nabla_{\bar{a}} , \nabla_{b} ] \epsilon  \ = \  & - \frac{1}{2} \gamma^{\bar{a}} \left[ E_{\bar{c}} \omega_{b \bar{a}}{}^{\bar{c}} - E_{b} \omega_{\bar{c} \bar{a} }{}^{\bar{c}} + \omega_{d \bar{a} }{}^{\bar{c}}\, \omega_{\bar{c} b} {}^{d} - \omega_{b \bar{a}}{}^{ \bar{d} }\, \omega_{\bar{c} \bar{d}}{}^{\bar{c}} \right] \epsilon \\
 = \ & - \frac{1}{2} \gamma^{\bar{a}} {\cal R}_{b \bar{a}} \epsilon \; ,
 \end{split}
 \ee
as claimed in (\ref{CurvIdent}).

Let us now turn to the second identity in (\ref{CurvIdent}) involving the scalar curvature. We compute
 \be
 \begin{split}
  \big( \gamma^{\bar{a}} \nabla_{\bar{a}} \gamma^{\bar{b}} \nabla_{\bar{b}} &- \nabla^{a} \nabla_{ a }\big) \epsilon
 \ = \ \Big( \gamma^{\bar{a}} E_{\bar{a}} - \frac{1}{4} \omega_{\bar{a} \bar{e} \bar{f}} \gamma^{\bar{a} \bar{e} \bar{f} } - \frac{1}{2} \omega_{\bar{a} \bar{e} }{}^{\bar{a}} \gamma^{\bar{e}}   \Big)   \Big( \gamma^{\bar{b}} E_{\bar{b}} - \frac{1}{4} \omega_{\bar{b} \bar{c} \bar{d}} \gamma^{\bar{b} \bar{c} \bar{d} } - \frac{1}{2} \omega_{\bar{b} \bar{c} }{}^{\bar{b}} \gamma^{\bar{c}}   \Big) \epsilon \\
  & - \Big(  E^{a} - \frac{1}{4} \omega^{a}{}_{\bar{e} \bar{f}} \gamma^{\bar{e} \bar{f}}  \Big)
  \Big(   E_{a} - \frac{1}{4} \omega_{a \bar{c} \bar{d}} \gamma^{\bar{c} \bar{d}}   \Big) \epsilon - \omega_{a}{}^{a b}  \Big(   E_{b} - \frac{1}{4} \omega_{ b \bar{c} \bar{d}} \gamma^{\bar{c} \bar{d}}   \Big) \epsilon \; .
 \end{split}
 \ee
As above, we work out the various powers $\gamma^{(p)}$ of gamma matrices separately, which here are non-trivial only for even $p$, and then
show that only the scalar part survives.
The $\gamma^{(6)}$ terms are easily seen to cancel,
 \be
\frac{1}{16} \omega_{\bar{a} \bar{e} \bar{f}  } \omega_{ \bar{b} \bar{c} \bar{d}   } \gamma^{ \bar{a} \bar{e} \bar{f} \bar{b} \bar{c} \bar{d} } \ = \  \frac{1}{16} \omega_{\bar{a} \bar{e} \bar{f}  } \omega_{ \bar{b} \bar{c} \bar{d}   } \gamma^{  \bar{b} \bar{c} \bar{d} \bar{a} \bar{e} \bar{f}  } \ = \ -  \frac{1}{16} \omega_{\bar{a} \bar{e} \bar{f}  } \omega_{ \bar{b} \bar{c} \bar{d}   } \gamma^{ \bar{a} \bar{e} \bar{f} \bar{b} \bar{c} \bar{d} } \ = \ 0 \; .
\ee
We have verified that the $\gamma^{(4)}$ and $\gamma^{(2)}$ structures cancel upon insertion of the explicit expressions for the determined connections,
which is a rather lengthy computation that we do not display here.  Let us finally turn to the scalar part (without gamma matrices). It reads
\be \label{wogamma}
 \Big[- E^{A} E_{A} + \omega_{a}{}^{b a } E_{b}  + \omega_{\bar{a}}{}^{ \bar{b} \bar{a} } E_{\bar{b}}   \Big] \epsilon + \frac{1}{2} \Big[ E_{\bar{a}} \omega_{\bar{b}}{}^{\bar{a} \bar{b}} + \frac{3}{4} \omega_{[\bar{a} \bar{b} \bar{c} ]}  \omega^{[\bar{a} \bar{b} \bar{c} ]}  - \frac{1}{2} \omega_{\bar{a}}{}^{\bar{c} \bar{a}} \omega_{\bar{b} \bar{c}}{}^{\bar{b}} + \frac{1}{4} \omega_{a \bar{b} \bar{c}} \omega^{a \bar{b} \bar{c}} \Big] \epsilon \;.
\ee
The terms in the first square bracket vanish. To see this, we write it out and insert the determined connections,
\be
\begin{split}
 \Big[ -& \sqrt{2}\big(E^{A} E_{A}{}^{M} \big) \partial_{M} - \sqrt{2}\big(\partial_{M} E_{b}{}^{ M}\big) E^{b} + 2 \big(E_{b}d\big) E^{b}
 - \sqrt{2} \big(\partial_{M} E_{\bar{b}}{}^{ M}\big) E^{ \bar{b} } + 2 \big(E_{\bar{b}}d) E^{\bar{b}}  \Big] \epsilon \\
\ = \ & \Big[- \sqrt{2}(E^{A} E_{A}{}^{M} ) \partial_{M} - \sqrt{2} \big(\partial_{M} E_{B}{}^{M}\big) E^{BN} \partial_{N} +2\big(E_{B}d\big)E^{B}\Big] \epsilon \ = \ 0 \; .
\end{split}
\ee
Here we used the strong constraint, which implies that $E_{B}d\,E^{B}\epsilon=0$.  Therefore, the only non-vanishing contribution is the
second bracket in (\ref{wogamma}), which is proportional to the scalar curvature (\ref{scalarcurv}).
We have thus shown
 \be
   \big( \gamma^{\bar{a}} \nabla_{\bar{a}} \gamma^{\bar{b}} \nabla_{\bar{b}} - \nabla^{a} \nabla_{ a }\big) \epsilon \ = \ -\frac{1}{4}{\cal R}\epsilon\;,
 \ee
as claimed in  (\ref{CurvIdent}).


\begin{thebibliography}{99}

\bibitem{Hull:2009mi}
  C.~Hull, B.~Zwiebach,
  ``Double Field Theory,''
  JHEP {\bf 0909}, 099 (2009).
  [arXiv:0904.4664 [hep-th]].

\bibitem{Hull:2009zb}
  C.~Hull, B.~Zwiebach,
  ``The Gauge algebra of double field theory and Courant brackets,''
  JHEP {\bf 0909}, 090 (2009).
  [arXiv:0908.1792 [hep-th]].

\bibitem{Hohm:2010jy}
  O.~Hohm, C.~Hull and B.~Zwiebach,
  ``Background independent action for double field theory,''
  JHEP {\bf 1007} (2010) 016
  [arXiv:1003.5027 [hep-th]].

\bibitem{Hohm:2010pp}
  O.~Hohm, C.~Hull and B.~Zwiebach,
  ``Generalized metric formulation of double field theory,''
  JHEP {\bf 1008} (2010) 008
  [arXiv:1006.4823 [hep-th]].

\bibitem{Siegel:1993th}
  W.~Siegel,
  ``Superspace duality in low-energy superstrings,''
  Phys.\ Rev.\  D {\bf 48}, 2826 (1993)
  [arXiv:hep-th/9305073],  
  ``Two vierbein formalism for string inspired axionic gravity,''
  Phys.\ Rev.\  D {\bf 47}, 5453 (1993)
  [arXiv:hep-th/9302036].

  \bibitem{Tseytlin:1990nb}
A.~A.~Tseytlin,
``Duality Symmetric Formulation Of String World Sheet Dynamics,''
Phys.\ Lett.\ B {\bf 242}, 163 (1990);
``Duality Symmetric Closed String Theory And Interacting Chiral Scalars,''
Nucl.\ Phys.\ B {\bf 350}, 395 (1991).

\bibitem{Hohm:2010xe}
  O.~Hohm, S.~K.~Kwak,
  ``Frame-like Geometry of Double Field Theory,''
  J.\ Phys.\ A {\bf A44}, 085404 (2011).
  [arXiv:1011.4101 [hep-th]],

\bibitem{Kwak:2010ew}
  S.~K.~Kwak,
  ``Invariances and Equations of Motion in Double Field Theory,''
  JHEP {\bf 1010} (2010) 047
  [arXiv:1008.2746 [hep-th]], \\
  O.~Hohm,
  ``T-duality versus Gauge Symmetry,''
  arXiv:1101.3484 [hep-th], \\
  O.~Hohm,
  ``On factorizations in perturbative quantum gravity,''
  JHEP {\bf 1104}, 103 (2011).
  [arXiv:1103.0032 [hep-th]], \\
  B.~Zwiebach,
  ``Double Field Theory, T-Duality, and Courant Brackets,''
  [arXiv:1109.1782 [hep-th]].


\bibitem{Hohm:2011ex}
  O.~Hohm, S.~K.~Kwak,
  ``Double Field Theory Formulation of Heterotic Strings,''
  JHEP {\bf 1106}, 096 (2011).
  [arXiv:1103.2136 [hep-th]].

\bibitem{Hohm:2011zr}
  O.~Hohm, S.~K.~Kwak, B.~Zwiebach,
  ``Unification of Type II Strings and T-duality,''
  Phys.\ Rev.\ Lett.\  {\bf 107}, 171603 (2011), 
    [arXiv:1106.5452 [hep-th]], 
  ``Double Field Theory of Type II Strings,''
  JHEP {\bf 1109}, 013 (2011), 
    [arXiv:1107.0008 [hep-th]].

\bibitem{Hohm:2011cp} 
  O.~Hohm and S.~K.~Kwak,
  ``Massive Type II in Double Field Theory,''
  JHEP {\bf 1111}, 086 (2011)
  [arXiv:1108.4937 [hep-th]].

\bibitem{Hillmann:2009ci}
  C.~Hillmann,
  ``Generalized E(7(7)) coset dynamics and D=11 supergravity,''
  JHEP {\bf 0903}, 135 (2009).
  [arXiv:0901.1581 [hep-th]].

\bibitem{Berman:2010is}
  D.~S.~Berman, M.~J.~Perry,
  ``Generalized Geometry and M theory,''
  JHEP {\bf 1106}, 074 (2011).
  [arXiv:1008.1763 [hep-th]], \\
  D.~S.~Berman, H.~Godazgar, M.~J.~Perry,
  ``SO(5,5) duality in M-theory and generalized geometry,''
  Phys.\ Lett.\  {\bf B700}, 65-67 (2011).
  [arXiv:1103.5733 [hep-th]], \\
  D.~S.~Berman, E.~T.~Musaev, M.~J.~Perry,
  ``Boundary Terms in Generalized Geometry and doubled field theory,''
  [arXiv:1110.3097 [hep-th]], \\
  D.~S.~Berman, H.~Godazgar, M.~Godazgar, M.~J.~Perry,
  ``The Local symmetries of M-theory and their formulation in generalised geometry,''
  [arXiv:1110.3930 [hep-th]], \\
  D.~S.~Berman, H.~Godazgar, M.~J.~Perry, P.~West,
  ``Duality Invariant Actions and Generalised Geometry,''
  [arXiv:1111.0459 [hep-th]].

\bibitem{West:2010ev}
  P.~West,
  ``$E_{11}$, generalised space-time and IIA string theory,''
  Phys.\ Lett.\  {\bf B696}, 403-409 (2011).
  [arXiv:1009.2624 [hep-th]], \\
  A.~Rocen, P.~West,
  ``E11, generalised space-time and IIA string theory: the R-R sector,''
  [arXiv:1012.2744 [hep-th]]. 
  
\bibitem{Jeon:2010rw}
  I.~Jeon, K.~Lee, J.~-H.~Park,
  ``Differential geometry with a projection: Application to double field theory,''
  JHEP {\bf 1104}, 014 (2011).
  [arXiv:1011.1324 [hep-th]].
  
\bibitem{Jeon:2011cn}
  I.~Jeon, K.~Lee, J.~-H.~Park,
  ``Stringy differential geometry, beyond Riemann,''
  Phys.\ Rev.\  {\bf D84}, 044022 (2011).
  [arXiv:1105.6294 [hep-th]]. 
  
\bibitem{Jeon:2011kp}
  I.~Jeon, K.~Lee and J.~-H.~Park,
  ``Double field formulation of Yang-Mills theory,''
  Phys.\ Lett.\ B {\bf 701} (2011) 260
  [arXiv:1102.0419 [hep-th]].
  
\bibitem{Jeon:2011vx}
  I.~Jeon, K.~Lee, J.~-H.~Park,
  ``Incorporation of fermions into double field theory,''
  JHEP {\bf 1111}, 025 (2011).
  [arXiv:1109.2035 [hep-th]].
  
\bibitem{Schulz:2011ye}
  M.~B.~Schulz,
  ``T-folds, doubled geometry, and the SU(2) WZW model,''
  [arXiv:1106.6291 [hep-th]].

\bibitem{Copland:2011yh}
  N.~B.~Copland,
  ``Connecting T-duality invariant theories,''
  Nucl.\ Phys.\  {\bf B854}, 575-591 (2012).
  [arXiv:1106.1888 [hep-th]], 
  ``A Double Sigma Model for Double Field Theory,''
  [arXiv:1111.1828 [hep-th]].

\bibitem{Thompson:2011uw}
  D.~C.~Thompson,
  ``Duality Invariance: From M-theory to Double Field Theory,''
  JHEP {\bf 1108}, 125 (2011).
  [arXiv:1106.4036 [hep-th]].

\bibitem{Albertsson:2011ux}
  C.~Albertsson, S.~-H.~Dai, P.~-W.~Kao, F.~-L.~Lin,
  ``Double Field Theory for Double D-branes,''
  JHEP {\bf 1109}, 025 (2011).
  [arXiv:1107.0876 [hep-th]].

\bibitem{Andriot:2011uh}
  D.~Andriot, M.~Larfors, D.~Lust, P.~Patalong,
  ``A ten-dimensional action for non-geometric fluxes,''
  JHEP {\bf 1109}, 134 (2011).
  [arXiv:1106.4015 [hep-th]], \\
  G.~Aldazabal, W.~Baron, D.~Marques, C.~Nunez,
  ``The effective action of Double Field Theory,''
  JHEP {\bf 1111}, 052 (2011).
  [arXiv:1109.0290 [hep-th]], \\
  D.~Geissbuhler,
  ``Double Field Theory and N=4 Gauged Supergravity,''
  [arXiv:1109.4280 [hep-th]].


\bibitem{deWit:1985iy}
  B.~de Wit, H.~Nicolai,
  ``Hidden Symmetry in d = 11 Supergravity,''
  Phys.\ Lett.\  {\bf B155}, 47 (1985),
  ``d = 11 Supergravity with local SU(8) invariance,''
  Nucl.\ Phys.\  {\bf B274}, 363 (1986).

\bibitem{Coimbra:2011nw}
  A.~Coimbra, C.~Strickland-Constable, D.~Waldram,
  ``Supergravity as Generalised Geometry I: Type II Theories,''
    [arXiv:1107.1733 [hep-th]].
    
\bibitem{Jeon:2011sq} 
  I.~Jeon, K.~Lee and J.~-H.~Park,
  ``Supersymmetric Double Field Theory: Stringy Reformulation of Supergravity,''
  arXiv:1112.0069 [hep-th].

\bibitem{Bergshoeff:1988nn}
  E.~Bergshoeff, M.~de Roo,
  ``Supersymmetric Chern-simons Terms In Ten-dimensions,''
  Phys.\ Lett.\  {\bf B218 } (1989)  210.

\bibitem{Bergshoeff1989}
E. A. Bergshoeff and M. de Roo,  "The quartic effective action of the heterotic string and supersymmetry", Nuclear Physics B 328 (1989) 439



\end{thebibliography}
\end{document}